\documentclass[useAMS,usenatbib]{mn2e}

\usepackage{times}
\usepackage{multirow}
\usepackage[T1]{fontenc} 
\usepackage{aecompl}

\usepackage{psfrag}
\usepackage{pspicture}
\usepackage{graphics}
\usepackage{graphicx}
\usepackage{amsmath}
\usepackage{color}

\title[Star-formation in merging galaxy clusters]{The rise and fall of star-formation in $\bf z\sim0.2$ merging galaxy clusters}
\author[A. Stroe et al.]{Andra Stroe$^{1}$\thanks{E-mail: astroe@strw.leidenuniv.nl}, David Sobral$^{1,2,3}$\thanks{VENI Fellow}, William Dawson$^{4}$, M. James Jee$^{5}$, Henk Hoekstra$^{1}$,
\newauthor David Wittman$^{5}$, Reinout J. van Weeren$^{6}$\thanks{NASA Einstein Postdoctoral Fellow}, Marcus Br\"uggen$^{7}$, Huub J. A. R\"ottgering$^{1}$\\
$^{1}$Leiden Observatory, Leiden University, P.O.\ Box 9513, NL-2300 RA Leiden, The Netherlands\\
$^{2}$Instituto de Astro\'{\i}sica e Ci\^{e}ncias do Espa\c{c}o, Universidade de Lisboa, Observat\'{o}rio Astron\'{o}mico de Lisboa, Tapada da Ajuda, 1359-018, Lisboa, Portugal\\ 
$^{3}$Centro de Astronomia e Astrof\'{\i}sica da Universidade de Lisboa, Observat\'{o}rio Astron\'{o}mico de Lisboa, Tapada da Ajuda, 1359-018, Lisboa, Portugal\\ 
$^{4}$Lawrence Livermore National Laboratory, P.O. Box 808 L-210, Livermore, CA, 94551, USA\\
$^{5}$Department of Physics, University of California, Davis, One Shields Avenue, Davis, CA 95616, USA\\
$^{6}$Harvard Smithsonian Center for Astrophysics (CfA - SAO), 60 Garden Street Cambridge, MA 02138\\
$^{7}$Hamburger Sternwarte, Universit\"at Hamburg, Gojenbergsweg 112, 21029 Hamburg, Germany
}

\begin{document}
\date{LLNL-JRNL-661314-DRAFT}
\maketitle

\begin{abstract}
CIZA J2242.8+5301 (`Sausage') and 1RXS J0603.3+4213 (`Toothbrush') are two low-redshift ($z\sim0.2$), massive ($\sim2\times10^{15}M_\odot$), post-core passage merging clusters, which host shock waves traced by diffuse radio emission. To study their star-formation properties, we uniformly survey the `Sausage' and `Toothbrush' clusters in broad and narrow band filters and select a sample of $201$ and $463$ line emitters, down to a rest-frame equivalent width ($13${\AA}). We robustly separate between H$\alpha$ and higher redshift emitters using a combination of optical multi-band (B, g, V, r, i, z) and spectroscopic data. We build H$\alpha$ luminosity functions for the entire cluster region, near the shock fronts, and away from the shock fronts and find striking differences between the two clusters. In the dynamically younger, $1$ Gyr old `Sausage' cluster we find numerous ($59$) H$\alpha$ emitters above a star-formation rate (SFR) of $0.17$ M$_{\sun}$ yr$^{-1}$ surprisingly located in close proximity to the shock fronts, embedded in very hot intra-cluster medium plasma. The SFR density for the cluster population is at least at the level of typical galaxies at $z\sim2$. Down to the same star-formation rate, the possibly dynamically more evolved `Toothbrush' cluster has only $9$ H$\alpha$ galaxies. The cluster H$\alpha$ galaxies fall on the SFR-stellar mass relation $z\sim0.2$ for the field. However, the `Sausage' cluster has an H$\alpha$ emitter density $>20$ times that of blank fields. If the shock passes through gas-rich cluster galaxies, the compressed gas could collapse into dense clouds and excite star-formation for a few $100$ Myr. This process ultimately leads to a rapid consumption of the molecular gas, accelerating the transformation of gas-rich field spirals into cluster S0s or ellipticals.
\end{abstract}

\begin{keywords}
galaxies: luminosity function, galaxies: evolution, shock waves, cosmology: observations, cosmology: large-scale structure of Universe, galaxies: clusters: individual: CIZA J2242.8+5301, 1RXS J0603.3+4213
\end{keywords}

\section{INTRODUCTION}\label{sec:intro}
Galaxy clusters grow by merging with other clusters and via accretion of galaxies \citep[e.g.][]{2002ASSL..272....1S}. Even at low redshifts ($z<0.5$), a significant population of galaxy clusters are undergoing mergers, with clear evidence from their disturbed intra-cluster medium (ICM) X-ray emission. Merging clusters are a unique probe of the interaction between dark matter, the ICM and the galaxies. They provide us with a way to test hierarchical structure formation, cosmic ray physics and galaxy evolution in dense environments. Major mergers have been argued to lead to increased turbulence within the ICM. In a number of cases merging clusters have been observed to produce travelling shock waves with Mach numbers ($M$) below $4$ \citep[e.g.][]{2014IJMPD..2330007B}. The shock fronts (re-)accelerate ICM electrons, which radiate synchrotron emission, observed in the radio as giant relics at cluster peripheries \citep{2014IJMPD..2330007B}.

\subsection{Star-forming galaxies in clusters}
The ICM interacts strongly with the cluster galaxies and is efficient in transforming the star-forming properties of member galaxies and/or maintaining them quenched \citep[e.g.][]{1978ApJ...226..559B, 1978ApJ...219...18B, 1980ApJ...236..351D}. The cluster galaxy population is dominated by passive, massive elliptical galaxies. The total galaxy number density in cluster environments is higher than in the field. Nevertheless, owing to a low fraction of blue, late-types within clusters, the number density of star-forming cluster galaxies is generally lower than in the field, \citep[e.g.][]{1980ApJ...236..351D, 2003MNRAS.346..601G}. For example, by using the H$\alpha$ emission line which traces recent ($<10$ Myr) star-formation, multiple authors have found that the number density of star-forming galaxies is $\sim50$ per cent lower than in blank fields \citep[e.g.,][]{2001ApJ...549..820C,2002MNRAS.335...10B,2004MNRAS.354.1103K}. Neutral hydrogen (HI) observations also show that cluster spirals contain significantly less HI gas than their field counterparts \citep[e.g.][]{1990AJ....100..604C}. 

Hence, dense cluster environments seem to suppress star-formation and probably lead to a morphological transformation from gas-rich spirals into gas-poor ellipticals. The deficit of star-forming galaxies in clusters is thought to occur through the process of ram pressure stripping \citep[e.g.][]{1972ApJ...176....1G, 2014arXiv1407.7527F}. Evidence of ram pressure stripping of the HI and H$\alpha$ gas in infalling cluster galaxies has been observed in the form of tails, knots and filaments \citep[e.g.][]{2001ApJ...563L..23G,2005A&A...437L..19O}. N-body, smooth particle hydrodynamical simulations by \citet{2012A&A...544A..54S}, in line with previous work by \citet{2003ApJ...596L..13B}, \citet{2008A&A...481..337K} and \citet{2009MNRAS.399.2221B}, show that relatively weak ram pressure can compress the inter-stellar medium of the galaxy and lead to an increase of star-formation. By contrast, high environmental densities and strong ram pressure can remove most of the gas from the host galaxy.

Other processes, such as galaxy harassment \citep{1996Natur.379..613M}, where galaxies are distorted by tidal forces, are also important. Tidal forces can be caused by the gravitational potential of the cluster or by encounters with other galaxies. The relative movement of galaxies as they fall into the cluster potential with respect to the ICM leads to a truncation of the outer galactic halo and disk. Simulations by \citet{1998ApJ...495..139M} indicate that galaxy harassment of small disk galaxies in clusters produces distorted spirals, often seen in $z\sim0.4$ clusters \citep[e.g.][]{1994ApJ...430..121C}, which evolve into spheroidal systems observed in local clusters.

\citet{1980ApJ...237..692L} proposed the process of galaxy strangulation as another means of transforming field spirals into cluster ellipticals and S0s. Gas from infalling galaxies escapes its host because of tidal forces created by the cluster potential well. With a limited supply of its main ingredient, the star-formation in the galaxy is effectively shut-down after over the course of a few Gyr.

\subsection{Merging clusters with shocks}
Even though the majority of galaxies in relaxed clusters are quenched, recent observations suggest that vigorous star-formation can be observed in merging clusters. By studying a sample of $z>0.3$ clusters, \citet{2014ApJ...781L..40E} found that gas within infalling galaxies is first shock compressed and then removed from the host galaxies. Therefore, if observed at the right time, gas-rich galaxies possibly shocked by infalling into the cluster or by the passage of a shock wave can exhibit high star-formation rates. 

There has been recent evidence that merging cluster processes such as increased turbulence and shocks affect the star formation properties of associated galaxies. \citet{2013A&A...557A..62P} find a population of quenched spirals at $3-4$ Mpc distances from the core of Abell 3921, which they attribute to shocks and cluster mergers. In the post-merger cluster Abell 2384, \citet{2014arXiv1408.0666P} find a significant population of disk galaxies in the cluster core, which is expected to be devoid of star-forming galaxies. \citet{2012ApJ...750L..23O} find three star-forming tails and filaments in galaxies nearby the X-ray shock front in Abell 2744. \citet{2003A&A...399..813F} and \citet{2004ApJ...601..805U} find a significant population of luminous, H$\alpha$-emitting, star-forming galaxies in the merging cluster Abell 521 \citep{2003A&A...399..813F, 2006A&A...446..417F}, which hosts a radio-detected shock front \citep{2008A&A...486..347G}. 

Galaxy formation simulations can be used to model the impact of mergers on galaxy properties.
For example, models of ram pressure stripping indicate that star-formation in galaxies within merging clusters can be quenched \citep[e.g.][]{2009A&A...499...87K}. More recently, hydrodynamical simulations of shocks passing through galaxies have reproduced star-forming tails trailing behind their parent galaxy \citep{roediger2014}. The star-formation persists for a few 100 Myr after the shock passage, in line with observations of star-forming tails in cluster infalling galaxies by \citet{2012ApJ...750L..23O}. 

\subsection{The `Sausage' and `Toothbrush' clusters}
\begin{figure*}
\begin{center}
\includegraphics[height=0.970\columnwidth]{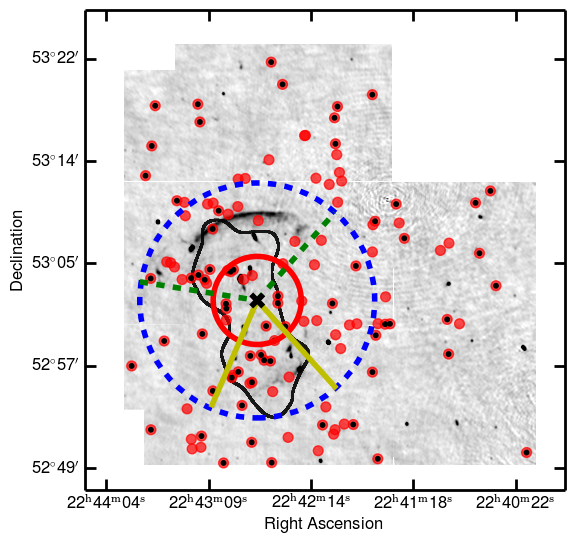}
\includegraphics[height=0.970\columnwidth]{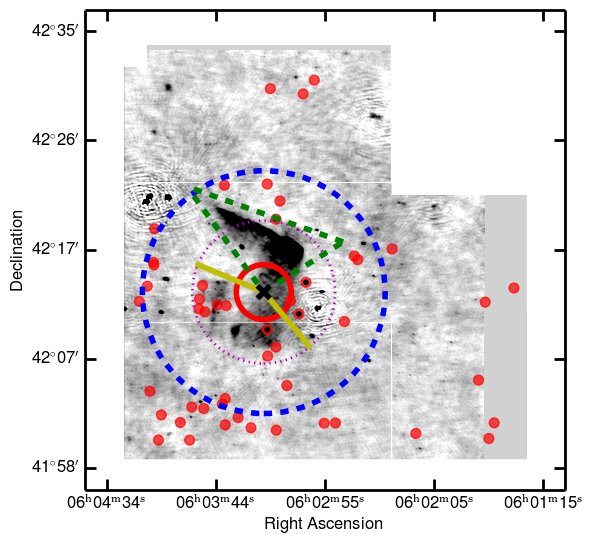}
\end{center}
\caption{The emitters for the `Sausage' field (left panel) and the `Toothbrush' field (right panel) in red circles and sources selected as H$\alpha$ emitters in the filled circles, overlaid on a Giant Meterwave Radio Telescope $323$~MHz radio images in grey intensity. The red circles with black dots at the centre represent spectroscopically confirmed H$\alpha$ emitters. The radio image is cut according to the NB FOV coverage, which was used for the selection of the emitters. 
The arc-sectors define the areas around the radio relics which were considered for producing relic luminosity functions. The cluster centres are defined to be at the location of the black crosses. \textit{Left}: `Sausage' cluster; The Northern relic was captured by a section between the solid red and the dashed blue circles, bound by the dashed green lines. The Southern relic area was defined between the solid red and the dashed blue circles and the solid yellow radii. The weak lensing area, enclosing a mass of about $1\times10^{15}M_\odot$, is marked by the solid black curve. The non-relic areas are defined as the arcsectors between the solid red and dashed blue circles and the dashed green and solid yellow lines. The entire cluster was assumed to occupy the volume defined by the dashed blue circle, which has an $\sim1.85$ Mpc radius. \textit{Right}: `Toothbrush' cluster. The Northern relic was captured by a section of the solid red circle, bound by the dashed green lines. The Southern relic area was defined between the solid red and dotted purple circles and the solid yellow radii. The non-relic areas are defined as the arcsectors between the solid red and dotted purple circles and the dashed green and solid yellow lines. The cluster was assumed to occupy the volume defined by the dashed blue circle of $\sim2.2$ Mpc radius.}
\label{fig:image}
\end{figure*}

To probe the effects of shocks in transforming cluster galaxies, we started an observing campaign of two major-merging, $z\sim0.2$ clusters hosting some of strongest radio shocks detected to date: CIZA J2242.8+5301 \citep[nicknamed the `Sausage',][]{2010Sci...330..347V} and 1RXS J0603.3+4213 \citep[nicknamed the `Toothbrush',][]{2012A&A...546A.124V}. The peculiar morphology of the relics explains the nickname of each cluster (see Fig.~\ref{fig:image}). The bright radio relics show clear signs of steepening and curving radio spectrum from the shock front into the downstream area, suggesting a scenario where the synchrotron electrons are shock accelerated and subsequently cool \citep{2010Sci...330..347V, 2012A&A...546A.124V, 2013A&A...555A.110S}. Nevertheless, the high-frequency, $16$ GHz observations of the `Sausage' relic suggests a more complicated scenario in which the electrons are injected and accelerated also in the downstream area \citep{2014MNRAS.441L..41S}. Both clusters are massive, X-ray luminous and present elongated X-ray morphologies suggesting a merger in the plane of the sky \citep{2013PASJ...65...16A, 2013MNRAS.429.2617O, 2013MNRAS.433..812O, 2014MNRAS.440.3416O}. They host two radio relics, which trace $M\sim2-4$ Mach number shocks. In both clusters, one relic is significantly larger and brighter than its counterpart, suggesting the merging sub-clusters were at close, but not $1:1$ mass ratio. A weak lensing analysis by \citet{Jee} indicates that the `Sausage' cluster is among the most massive clusters discovered to date, with a total mass exceeding of $M_{200}>2.5\times10^{15}M_{\odot}$. The northern ($M_\mathrm{N}=11.0^{+3.7}_{-3.2}\times10^{14}M_\odot$) and the southern sub-clusters \citep[$M_\mathrm{S}=9.8^{+3.8}_{-2.5}\times10^{14}M_\odot$][]{Jee} are very similar in mass.
\citet{dawson} derive velocity dispersion based mass estimates ($M_\mathrm{N}=16.1^{+4.6}_{-3.3}\times10^{14}M_\odot$ and 
$M_\mathrm{S}=13.0^{+4.0}_{-2.5}\times10^{14}M_\odot$), which are in agreement with the weak-lensing results.

On the basis of its X-ray luminosity, \citet{2012A&A...546A.124V} and \citet{2012MNRAS.425L..76B} conclude the `Toothbrush' is also a very massive cluster of about $1-2\times10^{15}M_\odot$.

Hydrodynamical simulations, radio spectral modelling and an analytical dynamics analysis suggest that the `Sausage' core-passage has happened $\sim1.0$ Gyr ago, at a relative speed of $\sim2000-2500$ km s$^{-1}$ \citep{dawson,2011MNRAS.418..230V,modelling}, making it a younger merger than the possibly $\sim2$ Gyr old `Toothbrush' merger \citep{2012MNRAS.425L..76B}. 

\subsection{This paper}
In this paper, we aim to study the star-formation properties of galaxies within the `Sausage' and the `Toothbrush' clusters. We derive star-formation rates and masses for the H$\alpha$ sample and build luminosity functions for different environments in and around the cluster volumes. 

By using narrow band observations, \citet{2014MNRAS.438.1377S} constrained the H$\alpha$ luminosity function (LF) of the cluster galaxies and found striking differences between the `Sausage' and the `Toothbrush' clusters. \citet{2014MNRAS.438.1377S} observed a notable enhancement in the normalisation of the LF around the relics within the `Sausage' cluster, where the travelling shock is expected to have passed $10-100$ Myr ago. By contrast, the `Toothbrush' cluster is almost devoid of line emitters, similar to a relaxed cluster. The relatively short time span when enhanced star-formation is seen in the `Sausage' cluster could explain the differences found between the two clusters, in line with simulations from \citet{roediger2014}.

In \citet{2014MNRAS.438.1377S}, lack of multi-band photometry prevented the galaxy-by-galaxy separation between H$\alpha$ emitters and higher redshift, lower rest-frame wavelength emission lines. Without such data, disentangling the drivers of the enhanced H$\alpha$ emission in the `Sausage' clusters is not possible, specifically the role of the post-merger timescales in shaping the H$\alpha$ properties of the galaxies. Instead, \citet{2014MNRAS.438.1377S} applied a statistical correction for the fraction of H$\alpha$ emitters from the total number of emitters, which was based on deep, narrow-band observations on the Cosmic Evolution Survey field \citep[COSMOS;][]{2008ApJS..175..128S}. 

In this paper, we present an optical multi-wavelength analysis of the `Sausage' and `Toothbrush' clusters. A combination of photometric and spectroscopic data from the Isaac Newton, William Herschel, Canada-France-Hawaii, Subaru and Keck telescopes spanning the entire optical spectrum through the B, V, g, r, i, z bands enables us to properly separate H$\alpha$ emitters from higher redshift emitters. Compared to \citet{2014MNRAS.438.1377S}, we are going $0.2-0.4$ magnitudes (mag) deeper in our detection band (i) and in the narrow band data, resulting in a larger sample of emitters, of which more than $50$ per cent have been followed up spectroscopically and confirmed as H$\alpha$ emitters. 

In \S\ref{sec:data} we present the reduction of the photometric and spectroscopic observations and the source extraction. \S\ref{sec:results} describes the emitter selection and separation, the luminosity functions and properties we derive for the galaxies. In \S\ref{sec:discussion} we compare our results with those for field galaxies and galaxies hosted in other clusters. We assume a flat $\Lambda$CDM cosmology with $H_{0}=70.5$~km~s$^{-1}$~Mpc$^{-1}$, matter density $\Omega_M=0.27$ and dark energy density $\Omega_{\Lambda}=0.73$ \citep{2009ApJS..180..306D}. We make use of Edward Wright's online cosmological calculator \citep{2006PASP..118.1711W}. One arcmin measures $0.191$~Mpc at $z=0.192$ (`Sausage'), while at $z=0.225$ (`Toothbrush') it corresponds to a physical size of $0.216$~Mpc. All images are in the J2000 coordinate system. Magnitudes are in the AB system. 

\begin{table}
\begin{center}
\caption{Filter properties: type (narrow band, NB, or broad band, BB), weighted central wavelength and full width at half maximum. The redshift range $z_{\mathrm{H}\alpha}$ for which the H$\alpha$ line is detected within the FWHM of the narrow band filters is also given.}
\begin{tabular}{l c c c c}
\hline
\hline
Filter & Type & $\lambda_\mathrm{c}$ ({\AA}) & FWHM ({\AA}) \\
\hline
NOVA782HA & NB & $7839.0$ & \phantom{0}$110$   \\
($z_{\mathrm{H}\alpha}=0.1865-0.2025$) & & &  \\
NOVA804HA & NB & $8038.5$ & \phantom{0}$110$  \\
($z_{\mathrm{H}\alpha}=0.2170-0.2330$) & &  \\
INT i & BB & $7746.0$ & $1519$  \\
WHT B & BB & $4332.7$ & $1065$  \\
Subaru g & BB & $4705.5$ & $1393$   \\
INT g & BB & $4857.3$ & $1290$   \\ 
WHT V & BB & $5488.1$ & \phantom{0}$990$    \\
INT V & BB & $5483.4$ & \phantom{0}$990$   \\
CFHT r   & BB & $6257.9$ & $1200$    \\
Subaru i & BB & $7676.0$ & $1555$   \\
WHT z & BB & $8720.9$  & -- \\
INT z & BB & $8749.3$ & -- \\
\hline
\end{tabular}
\label{tab:filters}
\end{center}
\end{table}

\section{DATA REDUCTION \& ANALYSIS}\label{sec:data}
We use a multitude of photometric and spectroscopic instruments mounted on a range of optical telescopes. We describe the data acquisition, reduction and processing below. Table~\ref{tab:filters} and Fig.~\ref{fig:transmittance} display the filter properties, while the observations and integration times can be found in Table~\ref{tab:observations}.

\subsection{Observations \& data processing}\label{sec:obs-reduction}

\subsubsection{Isaac Newton Telescope observations}
\label{sec:obs-reduction:INT}

The broad band (BB) i and narrow band (NB) NOVA782HA and NOVA804HA imaging data presented in \citet{2014MNRAS.438.1377S}, have been supplemented with new g, V, z, i and NB data taken in 4 photometric nights in October-November 2013 with the Wide Field Camera (WFC)\footnote{http://www.ing.iac.es/engineering/detectors/ultra\_wfc.htm} mounted on the Isaac Newton Telescope (INT, PI Stroe) \footnote{http://www.ing.iac.es/Astronomy/telescopes/int/}. The camera, a mosaic of four chips, has a $0.33$ arcsec pixel$^{-1}$ scale and a square field of view (FOV) of $34.2$ arcmin $\times$ $34.2$ arcmin, with the top north-western corner missing. Individual exposures of $200$ s for the BB and $600$ s for the NB were taken in 5 dithered positions to cover the chip gaps, under seeing conditions varying from $0.7$ to $2.0$ arcsec. A total of $\sim90$ ks and $\sim51$ ks were observed in the NB and $\sim13$ ks and $\sim12$ ks in the BB for the `Sausage' and `Toothbrush', respectively. For details see Table~\ref{tab:observations}.

Note that the NB filters have a full width at half maximum of $110$ {\AA}. They were designed to capture H$\alpha$ emission ($\lambda_\mathrm{restframe}=6562.8$\;{\AA}) at the redshift of the `Sausage' and `Toothbrush' clusters. See \citet{2014MNRAS.438.1377S} for further details.

\subsubsection{Canada France Hawaii Telescope observations}\label{sec:obs-reduction:CFHT}
Under OPTICON programme 13B055 (PI Stroe), service mode r-band images were taken using the Megacam imager\footnote{http://www.cfht.hawaii.edu/Instruments/Imaging/Megacam/} installed on the 3.6-m Canada-France-Hawaii Telescope (CFHT)\footnote{http://www.cfht.hawaii.edu/}, under excellent seeing conditions ($<0.8$ arcsec), between July and December 2013. The 36-chip camera has a $\sim1$ deg$^2$ FOV, with a $0.187$ arcsec pixel$^{-1}$. To obtain a contiguous FOV coverage, $600$ s exposures were taken in two dither positions spaced at $15$ arcmin. $18$ ks were observed in the `Sausage' field and $5.4$ ks in the `Toothbrush'.

\subsubsection{Subaru observations}\label{sec:obs-reduction:Subaru}
Images in the g and i band (PI Wittman) were taken with Subaru's\footnote{http://www.naoj.org} Prime Focus Camera (Suprime-Cam)\footnote{http://www.naoj.org/Observing/Instruments/SCam/index.html}, a 10-chip mosaic with a $34$ arcmin $\times$ $27$ arcmin FOV and $0.2$ arsec pixel scale. For the full details of the observations we refer the reader to \citet{dawson} and \citet{Jee}.

\begin{figure}
\begin{center}
\includegraphics[trim =0cm 0cm 0cm 0cm, width=0.46\textwidth]{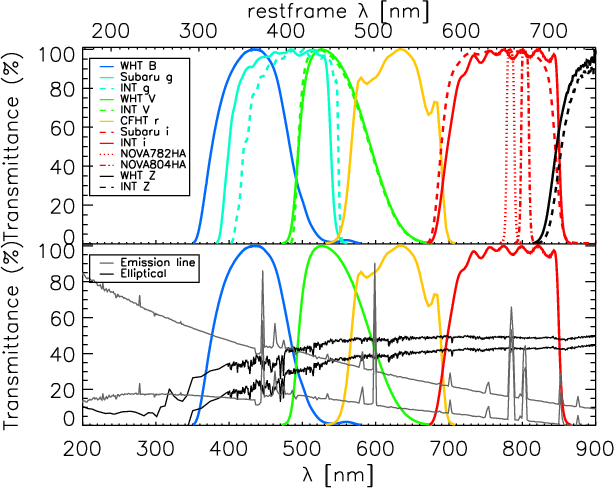}
\end{center}
\caption{Top panel: Normalised transmittance profiles for the filters used in the analysis. 
The top x-axis marks the restframe wavelength coverage of the filters, assuming a redshift of $0.2$. 
%The solid red line marks the profile of the INT i band filter used as our main BB filter. The dotted and the dash-dotted red lines defines the NOVA782HA and NOVA804HA filters, respectively. The dashed red line is the Subaru i band filter. The black lines are z band from INT and WHT. The yellow line is CFHT r, while the green is V from WHT and INT. WHT B is marked in the dark blue solid line. The cyan lines represent the INT and Subaru g bands. 
Bottom panel: Emission spectra for passive and emission line galaxies redshifted to $z=0.2$, at arbitrary normalisation, taken from the PEGASE 2.0 template set \citep{2006A&A...449..951G}. Note how the H$\alpha$ line falls within the corresponding narrow-band filter NOVA782HA, as expected. Both intrinsic and dust-attenuated spectra are shown. Notice the effect of Galactic dust extinction, especially in the blue-side of the spectrum.}
\label{fig:transmittance}
\end{figure}

\subsubsection{William Herschel Telescope Imaging}\label{sec:obs-reduction:WHT}
B, V and z band data were taken on 1-3 November 2013 with the newly-commissioned Prime Focus Imaging Platform (PFIP)\footnote{http://www.ing.iac.es/Astronomy/instruments/pfip/index.html} on the 4.2-m William Herschel Telescope (WHT, PI Stroe)\footnote{http://www.ing.iac.es/Astronomy/telescopes/wht/}. The single-chip camera has a pixel scale of $0.27$ arcsec with a FOV of $18.0$ arcmin $\times$ $18$ arcmin. We took a series of $600$-s exposures in three pointings, jittered over five positions, to roughly cover the FOV of INT's WFC. The seeing varied between $0.7$ and $2.0$ arcsec. See table \ref{tab:observations} for the integration times.

\subsection{Spectroscopy}
We have also obtained spectroscopic data using two instruments. Spectra of a sample of $27$ emission line, star-forming galaxies in the `Sausage' and the `Toothbrush' clusters were taken with the DEep Imaging Multi-Object Spectrograph \citep[DEIMOS,][]{2003SPIE.4841.1657F}\footnote{http://www2.keck.hawaii.edu/inst/deimos/} mounted at the Nasmyth focus of the Keck II telescope at the W. M. Keck Observatory (PI Wittman)\footnote{http://www.keckobservatory.org/}. These observations are described in detail in \citet{dawson}.

Spectra of line emitters within the `Sausage' cluster roughly covering the $5500-8500${\AA} range were obtained on 2 and 3 July 2014 with the multi-object, wide-field AutoFib2 (AF2)\footnote{http://www.ing.iac.es/Astronomy/instruments/af2/index.html} fibre spectrograph mounted at the prime focus of WHT (PI Stroe). A series of 30 min exposures in two configurations allowed us to observe $\sim100$ sources within the cluster and its northern periphery. Six line emitters in this sample were also targeted with DEIMOS. The data were analysed similarly to the method described in \citet{dawson}. In short, the fibre traces on the CCD was corrected for curvature using the lamp flats. After bias-subtracting, flattening and sky-background subtracting the data, the spectra were extracted and wavelength-calibrated using the lamp flats and sky lines. Full details can be found in Sobral et al. (in prep) and \citet{dawson}.

Note that the observing strategies for the DEIMOS and AF2 data were fundamentally different. The DEIMOS observations, tailored for a redshift analysis were mainly targeting the red sequence galaxies. The AF2 observations, however, specifically targeted emission line galaxies presented in this paper.

\begin{table*}
\begin{center}
\caption{Details of the observations: filters, total integration times, effective integration times after removing bad frames and observing dates.}
\begin{tabular}{l l c c c c l}
\hline
\hline
Field & RA & DEC & Filter & Int. Time (ks) & Eff. Time (ks) & Dates \\
\hline
\multirow{9}{*}{`Sausage'} & \multirow{9}{*}{$22^{h}42^{m}50^{s}$} & \multirow{9}{*}{$53^{\circ}06'30''$} & NOVA782HA & $78.9$ & $47.4$ & October 13--15, 20--22 2012;  November 1--6 2013 \\
 & & & INT i & $13.2$ & \phantom{0}$9.0$ & October 15, 20--22, 2012;  November 1--6, 2013  \\
 & & & WHT B & $12.0$ & \phantom{0}$9.6$ & November 1,3 2013  \\
 & & & Subaru g & \phantom{0}$3.3$ & \phantom{0}$3.3$ & July 13, 2013  \\
 & & & WHT V &  \phantom{0}$9.0$ & \phantom{0}$9.0$ & November 1,3 2013  \\
 & & & CFHT r & $18.0$ & $18.0$ & July 3, 5--8, 11, 12 2013  \\
 & & & Subaru i & \phantom{0}$3.3$ & \phantom{0}$3.3$ & July 13, 2013  \\
 & & & WHT z &  $10.8$ & \phantom{0}$9.0$ & November 1,3 2013  \\
 & & & INT z &  \phantom{0}$1.0$ & \phantom{0}$1.0$ & November 1--6 2013  \\
\hline
\multirow{11}{*}{`Toothbrush'} & \multirow{11}{*}{$06^{h}03^{m}30^{s}$} & \multirow{11}{*}{$42^{\circ}17'30''$} & NOVA804HA & $51.0$ & $37.2$ & October 13--16, 20, 21 2012; November 1--6 2013 \\
 & & & INT i &$11.8$ & $11.8$ & October 15, 16, 21 2012; November 1--6 2013 \\
 & & & WHT B & \phantom{0}$12.0$ & \phantom{0}$9.6$ & November 1,3 2013  \\
 & & & INT g & \phantom{0}$6.0$ & \phantom{0}$6.0$ & November 1--6 2013  \\
 & & & WHT V &  \phantom{0}$9.0$ & \phantom{0}$9.0$ & November 1,3 2013  \\
 & & & INT V & \phantom{0}$2.0$ & \phantom{0}$2.0$ & November 1--6 2013  \\
 & & & CFHT r & \phantom{0}$5.4$ & \phantom{0}$5.4$ & December 4, 5 2013  \\
 & & & WHT z &  \phantom{0}$10.8$ & \phantom{0}$9.0$ & November 1, 3 2013  \\
 & & & INT z &  \phantom{0}$5.0$ & \phantom{0}$5.0$ & November 1--6 2013  \\
\hline
\end{tabular}
\label{tab:observations}
\end{center}
\end{table*}

\begin{table*}
\begin{center}
\caption{Observed $1\sigma$ error and $3\sigma$ limiting magnitudes (measured in $5$ arcsec apertures) for the `Sausage' and `Toothbrush' observations, uncorrected for the effects of Galactic dust attenuation. The extinction $A_\lambda$ ranges for that filter are also given.}
\begin{tabular}{l c c c c c c c c c}
\hline
\hline
`Sausage' & INT NB & INT i & INT z & WHT B & WHT V & WHT z & Subaru g & Subaru i & CFHT r \\
\hline
$1\sigma$ & 21.7 & 21.8 & 19.4 & 24.0 & 23.1 & 21.5 & 24.2 & 23.7 & 23.4 \\
$3\sigma$ & 20.5 & 20.7 & 18.2 & 22.8 & 21.9 & 20.2 & 23.1 & 22.5 & 22.2 \\
$A_\lambda$ & $0.6-1.0$ & $0.6-1.0$ & $0.5-0.8$ & $1.3-2.1$ & $1.0-1.6$ & $0.5-0.8$ & $1.2-1.9$ & $0.6-1.0$ & $0.8-1.3$ \\
\hline
\hline
`Toothbrush' & INT NB & INT i & INT g & INT V & INT z & WHT B & WHT V & WHT z & CFHT r \\
\hline
%$1\sigma$ & 21.5 & 22.5 & 23.5 & 22.8 & 20.8 & 24.0 & 23.6 & 21.5 & 23.8 \\
%$3\sigma$ & 20.3 & 21.3 & 22.3 & 22.3 & 19.6 & 22.8 & 22.4 & 20.3 & 22.6 \\
$1\sigma$ & 21.9 & 22.5 & 23.5 & 22.8 & 20.8 & 24.0 & 23.6 & 21.5 & 23.8 \\
$3\sigma$ & 20.7 & 21.3 & 22.3 & 22.3 & 19.6 & 22.8 & 22.4 & 20.3 & 22.6 \\
$A_\lambda$ & $0.32-0.43$ & $0.34-0.46$ & $0.27-0.37$ & $0.65-0.88$ & $0.57-0.76$ & $0.74-1.00$ & $0.56-0.76$ & $0.27-0.37$ & $0.47-0.68$ \\
\hline
\end{tabular}
\label{tab:limmag}
\end{center}
\end{table*}

\subsection{Photometric reduction and source extraction}
\label{sec:obs-reduction:red}
We reduced the BB and NB optical photometry from the INT, WHT, Subaru and CFHT using the standard approach for reducing imaging data, implemented in our in-house \textsc{python}-based pipeline. We rejected data affected by cloud extinction, pointing, focussing, read-out issues and very poor seeing ($>2$ arcsec). The data for each chip and each filter in the WFC (INT), Suprime-Cam (Subaru) and Megacam (CFHT) CCD mosaics were processed independently. Note that PFIP (WHT) imager contains a single CCD. 

The sky flats for each filter on each instrument were median-combined to obtain a `master-flat'. A `master-bias' for each night of observing was obtained by median-combining biases taken with each instrument. The individual exposures were bias-subtracted and sky-flattened to remove electronic camera noise, shadowing effect and normalise for the pixel quantum efficiency. Science exposure pixels that deviated by more than $3\sigma$ from the local median were blanked as non-responsive, hot or as cosmic rays. We additionally normalised the WHT and INT i and z bands by a `super-flat', obtained by combining science frames with masked sources. This step is necessary to remove the effects of significant thin-film interference (`fringing') for images taken in the red and near infra-red part of the spectrum.

We used recursive rounds of {\sc SCAMP} \citep{2006ASPC..351..112B} to find astrometric solutions for the processed exposures. $0.2-0.3$ arcsec root-mean-square (rms) accuracy per object was obtained by comparing source positions with 2MASS astrometry \citep{2006AJ....131.1163S}. The exposures were normalised to the same zero-point (ZP) by referencing to the closest photometric band measured in the fourth United States Naval Observatory (USNO) CCD Astrograph Catalog \citep[UCAC4;][]{2013AJ....145...44Z}. The fully-processed science exposures for each filter and each instrument were median-combined to obtain final stacked images using {\sc SWarp} \citep{2002ASPC..281..228B}. 

The Sloan Digital Sky Survey (SDSS) does not cover our fields. We therefore used the USNO-B1.0 catalogue \citep{2003AJ....125..984M} to derive the photometric calibration, as outlined in \citet{2014MNRAS.438.1377S}. The USNO-B1.0 magnitudes were converted to Johnson system B, V, z and Sloan system g, r, i based on relations derived from SDSS Data Release 7 \citep[SDSS DR7][]{2009ApJS..182..543A, 2009yCat.2294....0A} on a $9$ deg$^2$ field with overlapping SDSS DR7 and USNO-B1.0 coverage. Bright, but not saturated stars in our fields were matched to magnitudes of USNO-B1.0 sources, converted to the equivalent filter, in order to obtain the photometric ZP. Given the high number of sources matched an accuracy of $\sim0.05$ mag was attained in the calculation of the ZP. The calibration was performed independently for the four WFC chips (INT) and the three PFIP (WHT) pointings. 

We extracted sources using {\sc SExtractor} \citep{1996A&AS..117..393B}, measuring magnitudes in $5$ arcsec apertures, corresponding to a physical diameter of $\sim17$ kpc at the redshift of the clusters. This aperture ensures that we encompass the full disk of the galaxies. Subsequently, the magnitudes were corrected for dust absorption by the Milky Way, using the reddening values from \citet{2011ApJ...737..103S}, interpolated to the effective wavelength of each of our filters. The clusters are located at low Galactic latitude and suffer from significant dust extinction ($A_\lambda$) which varies across the relatively large FOV of our observations (see Figs.~\ref{fig:dust:sausage} and \ref{fig:dust:toothbrush}). If uncorrected for, the dust  extinction can shift galaxy B-V colours by up to $0.5$ magnitudes and B-z up to $1.5$ mag. 

We used the RMS noise reported by {\sc SExtractor} to calculate the $1\sigma$ and $3\sigma$ limiting magnitudes for our observations. Note that the Galactic dust extinction is substantial for our field and rises steeply in the blue side of the spectrum (see Table~\ref{tab:limmag}). 

\section{METHODS AND RESULTS}
\label{sec:results}

\subsection{Narrow band emitter selection}
\label{sec:obs-reduction:NBselection}
In order to select line-emitting candidates, we study the excess of the NB emission as compared to the BB continuum. If an emission line is present, the source will have a significant BB-NB colour excess. We use the same approach described in detail in \citet{2014MNRAS.438.1377S}, based on the methods of \citet{1995MNRAS.273..513B} and \citet{2009MNRAS.398...75S, 2012MNRAS.420.1926S}. We refer the interested readers to those papers for the details of the selection criteria. 

The different effective central wavelengths of the NB and BB filters (see Table~\ref{tab:filters}, $\sim100${\AA} for the `Sausage' and $\sim300${\AA} for the `Toothbrush') cause systematic BB-NB colour offsets. A constant offset was sufficient to correct the excess in the `Sausage' field, as there was no dependence of NB-BB excess on the NB magnitude. The NOVA804HA filter peaks $300${\AA} redder than the INT i BB filter. Therefore, the i and z magnitudes were used to correct for the colour offsets that vary with NB magnitude. For sources without a z band magnitude, a statistical correction was applied based on the average z-i colour. Note this is a significant improvement, greatly reducing scatter compared to \citet{2014MNRAS.438.1377S},  where z band data was not available.

The selection of emitters is performed anew, since our new NB and BB data are deeper by $0.2-0.4$ mag than in \citet{2014MNRAS.438.1377S}, with a better colour correction, allowing us to probe fainter and lower equivalent width (EW) emitters. In short, to be selected as a line emitter, a source must fulfil three criteria \citep[for details, see][]{2014MNRAS.438.1377S}:
\begin{itemize}
\item Significant excess ($\Sigma$) narrow band (NB) emission with respect to the broad band (BB, $\Sigma>3$), based on the scatter of the faint-end of the NB magnitudes. This criterion rejects faint, low signal-to-noise sources from entering the emitter catalogue.
\item A NB minus BB colour cut, intended to retain sources with an intrinsic emission line equivalent width (EW) higher than $13${\AA} (assuming the sources are at $z\sim0.2$). This ensures that we select sources with strong spectral features and remove stars or sources without an emission line, but which have steep continuum. The EW cut value was chosen to reflect the $3\sigma$ scatter of the BB-NB excess around $0$, for bright, but not saturated NB magnitudes.
\item Visual inspection to remove saturated stars, double stars and false positives at the edge of chips. These types of sources can mimic line emitters. 
\end{itemize}

\begin{figure*}
\begin{center}
\includegraphics[trim=0cm 0cm 0cm 0cm, width=0.495\textwidth]{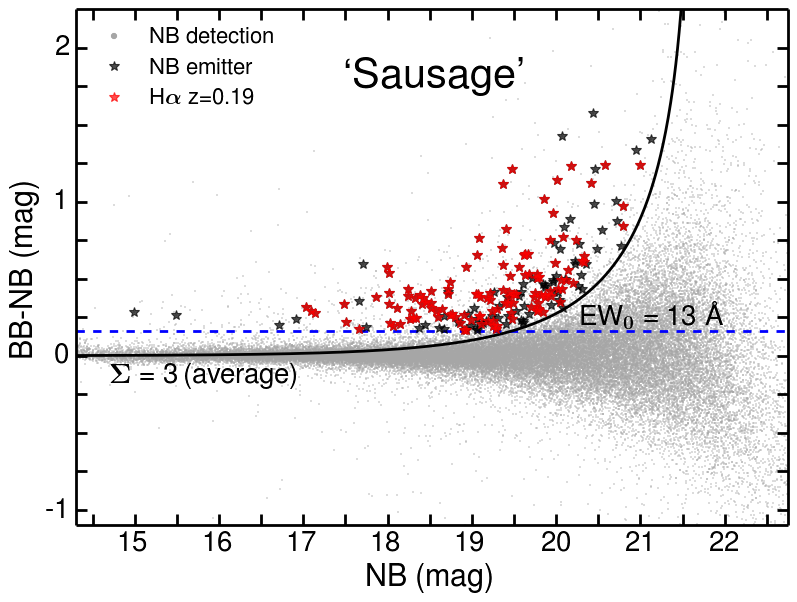}
\includegraphics[trim=0cm 0cm 0cm 0cm, width=0.495\textwidth]{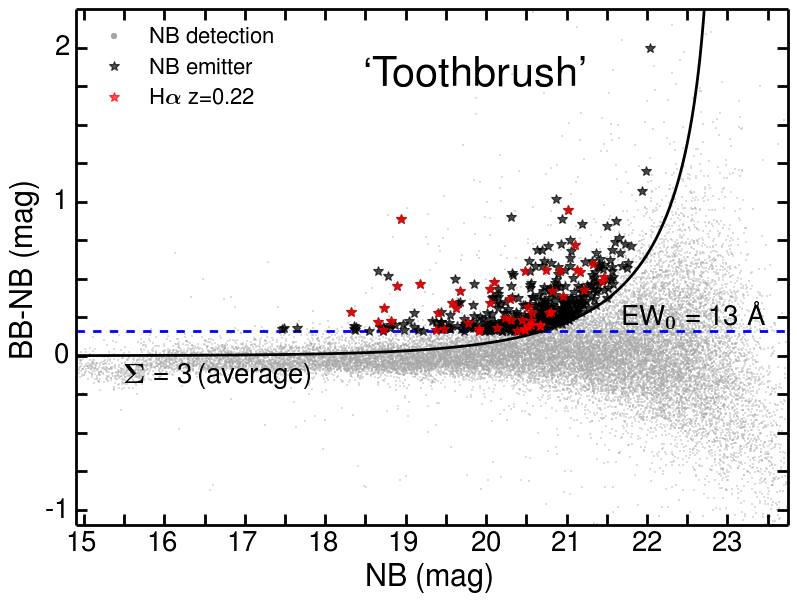}
\end{center}
\caption{BB minus NB as function of NB magnitude for the `Sausage' (left) and `Toothbrush' (right) fields. The INT i band was used for BB subtraction. The blue dashed, horizontal line represents the limiting rest-frame EW, while the curve marks the $3\Sigma$ colour significance limit for choosing sources as NB emitters (masked in the black stars). The sources selected as H$\alpha$ emitters at $z\approx0.2$ (according to Fig.~\ref{fig:colcol}) are shown in the red stars.}
\label{fig:colmag}
\end{figure*}

Our sample consists of $201$ emitters for the `Sausage' field ($0.020$ sources per kpc$^2$, down to a line flux of $5.2\times10^{-16}$ erg s$^{-1}$ cm$^{-2}$) and $463$ for the `Toothbrush' field ($0.036$ sources per kpc$^2$, see also Table~\ref{tab:emitters}, down to a line flux of $9.8\times10^{-17}$ erg s$^{-1}$ cm$^{-2}$). This is a substantial increase compared to \citet{2014MNRAS.438.1377S}, where $181$ emitters were found for the `Sausage' field and $141$ in the `Toothbrush' FOV.

\subsection{Identifying H$\alpha$ emitters among the line emitters}\label{sec:results:Halpha}
Our NB emitter population is composed of a sample of H$\alpha$ sources at $z\sim0.2$, together with other strong, higher-redshift line-emitters that fall within the passband of our NB filter. Nevertheless, given the moderate depth of our NB survey, we expect our emitter population to have a higher fraction of H$\alpha$ emitters compared to what is measured from deep surveys such as COSMOS, which are saturated at bright luminosities because of long individual integration times. Emission line sources strong enough to potentially be detected are: H$\beta$ ($\lambda_\mathrm{rest}=4861$\;{\AA}), [O{\sc iii}]$\lambda\lambda4959,5007$ emitters at $z\sim0.61-0.65$ and [O{\sc ii}] ($\lambda_\mathrm{rest}=3727$\;{\AA}) emitters at $z\sim1.15$. We might also be contaminated by $z\sim0.8$ 4000\;{\AA} break galaxies.

In order to differentiate between these emitter populations, we use colour-colour separation \citep[e.g.][]{2008ApJS..175..128S,2013MNRAS.428.1128S}, in combination with spectroscopic and photometric redshifts. In order to do so, we fully exploit the wealth of multiband photometry we have acquired. We use our INT i catalogue as our main detection catalogue (used for subtraction of the continuum, with the INT NB data). The Subaru i band catalogue, albeit deeper than the INT i, does not have perfectly matching FOV coverage to our NB observations. We note that all our emitters have a detection in the BB and NB. We further use all the data available from the other bands ranging from B up to z band. Given the different depths and FOV coverage of these data, not all emitters have detections in all seven ancillary bands. About $40$ per cent of sources have detections in all bands and another $\sim45$ per cent miss a detection in one single band. The remaining $\sim15$ per cent of sources lack $2$ or more bands.

\subsubsection{Colour-colour separation}
\begin{figure*}
\begin{center}
\includegraphics[trim=0cm 0cm 0cm 0cm, width=0.7895\textwidth]{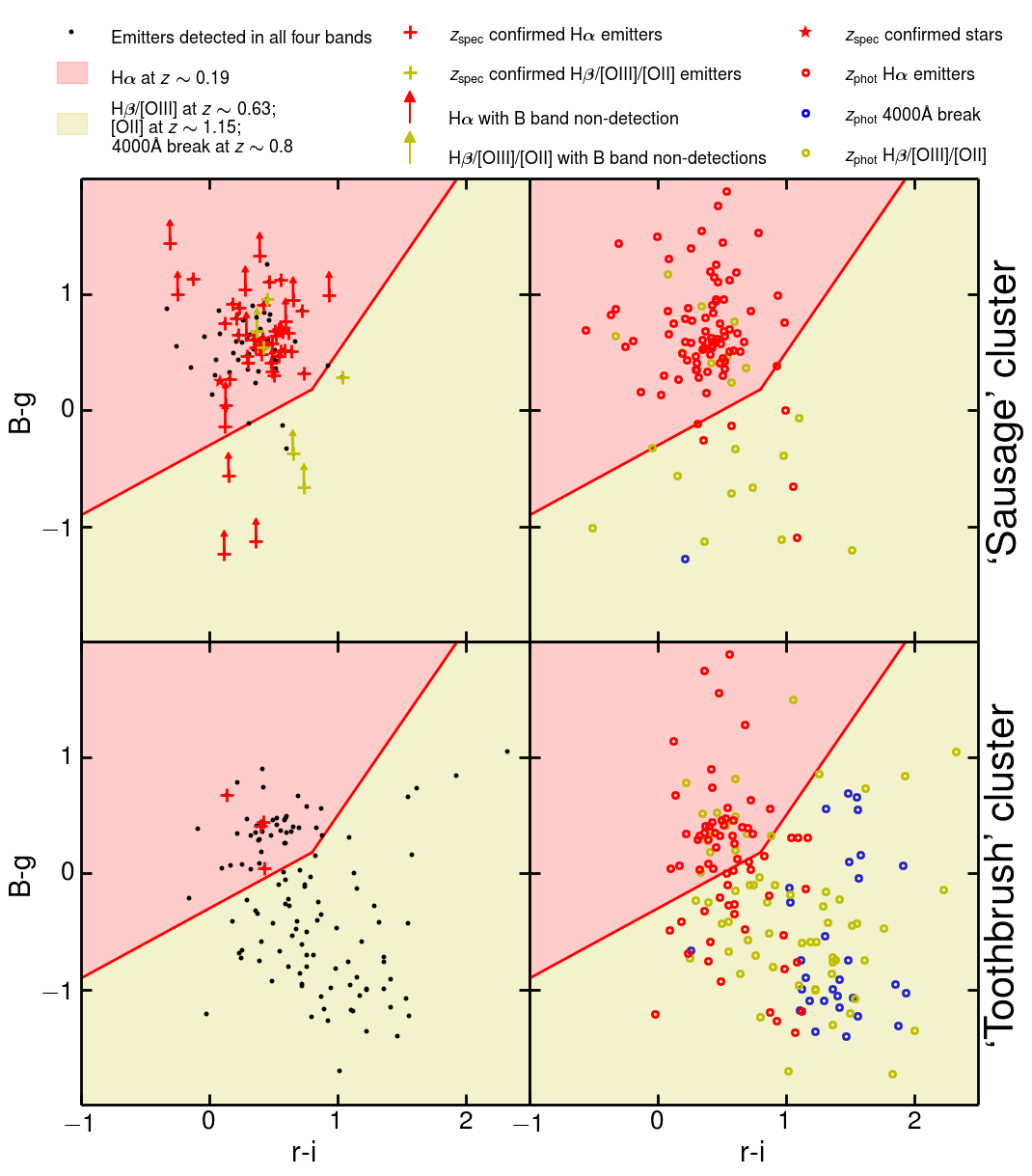}
\end{center}
\caption{Colour-colour plot of `Sausage' (top) plot and `Toothbrush' (bottom plot) emitters, selected according to the criteria described in Fig.~\ref{fig:colmag}. $B-g$ versus $r-i$ colours are plotted. H$\alpha$ emitters are expected to lie above and to the left of the thick red lines, in the red shaded areas. H$\beta$, [O{\sc iii}], [O{\sc ii}] emitters and $4000${\AA} break galaxies lie in the yellow shaded area. \textit{Left panels}: Sources with clear detections are plotted with black dots. Spectroscopically confirmed H$\alpha$ emitters are plotted in red crosses. \textit{Right panel}: Sources with photometric redshifts consistent with being H$\alpha$ emitters at $z\sim0.2$ are plotted with red circles. Possible $4000${\AA} break galaxies are plotted in blue. Sources with photo-z's around $0.63$ (between $0.5$ and $0.7$) and $1.15$ (between $1.0$ and $1.2$) are selected as H$\beta$, [O{\sc iii}] and [O{\sc ii}] emitters (plotted in yellow).}
\label{fig:colcol}
\end{figure*}

We base our colour-colour selection on the $z=0.24$ H$\alpha$ emitters selected in COSMOS \citep{2007ApJS..172...99C, 2009ApJ...690.1236I} from Subaru NB NB816 imaging \citep{2008ApJS..175..128S}. Since COSMOS goes to much fainter magnitudes than our data, we select only emitters with line emission greater than $2.5\times10^{-16}$ erg s$^{-1}$ cm$^{-2}$ to match the range observed in our survey. We explore possible colour-colour selections which best separate the low-redshift H$\alpha$ emitters from the higher-redshift interlopers. We use the photometric and spectroscopic redshifts available for COSMOS to test how many emitters are correctly classified by each colour-colour selection. We adopt the $\mathrm{B}-\mathrm{g}$ versus $\mathrm{r}-\mathrm{i}$ colour-colour plane as best discriminator between the low and high redshift emitters (see Fig.~\ref{fig:colcol}). Sources are selected as potential H$\alpha$ emitters if they simultaneously fulfil the two colour requirements:
\begin{align}
\label{eq:colcol}
(\mathrm{B}-\mathrm{g} ) &> (0.6(\mathrm{r}-\mathrm{i})-0.3) \\
(\mathrm{B}-\mathrm{g} ) &> (1.6(\mathrm{r}-\mathrm{i})-1.1)
\end{align} 
These separation lines are marked in Fig.~\ref{fig:colcol} by thick red lines. 

\subsubsection{Spectroscopic and photometric redshifts}
The redshift of the sources was found by measuring the position of the brightest emission lines (e.g. H$\alpha$, H$\beta$, [O{\sc iii}] [N{\sc ii}] and [S{\sc iii}]) in each spectrum. For the `Sausage' cluster, based on Keck spectra, there are $23$ emitters which have been spectroscopically confirmed as H$\alpha$ line galaxies at $z\sim0.2$ \citep[Sobral et al. in prep,][]{dawson}. Forty-eight H$\alpha$ sources are detected among the WHT spectra (see Sobral et al. in prep for details). 
Six emitters were targeted by both surveys and are detected as H$\alpha$ in both data sets, confirming the robustness of our analysis. Therefore, we have $65$ spectroscopically confirmed H$\alpha$ sources out of a sample of $201$ emitters. We confirm $8$ H$\beta$/[O{\sc iii}] emitters at $z\sim0.6$, 1 [O{\sc ii}] emitter at $z\sim1.1$ and 1 passive galaxy at $z\sim0.8$. Hence out of $75$ emitters with spectra, $65$ are H$\alpha$ ($87$ per cent) and the rest are higher redshift emitters. This low number of non-H$\alpha$ sources is partly driven by our selection of bright emitters (with a mean $5$ arcsec i band magnitude of $19.6$ compared to $19.9$ for the entire emitter population) for spectroscopic follow up, which have very high chances of being low rather than high-redshift. In addition, the Keck spectra confirm only $3$ stars which contaminate our emitter population and we note that we have not done any star rejection due to the lack of near infra-red data. We exclude these sources from our H$\alpha$ catalogue.

For the `Toothbrush' cluster, we have only $4$ spectroscopically confirmed H$\alpha$ emitters from Keck data. Only these $4$ emitters were serendipitously targeted as the survey was targeting passive members.

\begin{table*}
\begin{center}
\caption{Statistics for the emitters detected in our `Sausage' and `Toothbrush' surveys. A non-detection in a particular band can be caused by lack of FOV coverage or limited depth. For emitters with detections in all bands or with significant upper limits, a colour-colour selection criterion could be applied to select H$\alpha$ emitters. Additionally, spectroscopic or photometric redshifts were used to select potential H$\alpha$ sources. Note that even though the sample of emitters is larger for the `Toothbrush' FOV, the number of H$\alpha$ emitters is smaller, reinforcing the effect seen in Fig.~\ref{fig:photoz}, where the photometric redshift distribution is dominated by high-z emitters.}
\begin{tabular}{l l r r r r r r r r r r}
\hline
\hline
Field & Emitters & \multicolumn{7}{}{*}{Detected in} & Not detected &  $z_\mathrm{spec}$ & H$\alpha$ selected \\ 
     &    & B, g, r, i  &  B, g ,i     & B, r, i    & g, r, i     & B & g & r    & in B, g, r & selected &  \\
\hline
`Sausage' & $201$   & $79$  		&	$0$	   & $2$    & $95$    & $0$ & $0$  & $25$ & $0$  & $65$ & $124$ \\
`Toothbrush' & $463$   & $120$  	&	$9$	   & $51$    & $114$  & $3$ & $6$  & $141$ & $19$  & $4$ & $50$ \\
\hline
\end{tabular}
\label{tab:emitters}
\end{center}
\end{table*}

\begin{figure}
\begin{center}
\includegraphics[trim=0cm 0cm 0cm 0cm, width=0.495\textwidth]{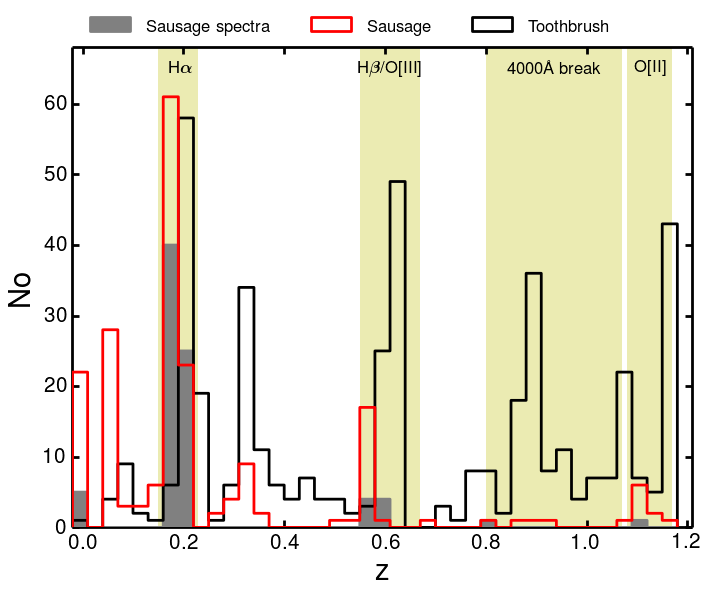}
\end{center}
\caption{Redshift distribution of emitters in the `Sausage' and `Toothbrush' FOV, as reported photometrically by \textsc{eazy}. Spectroscopic redshifts obtained from Keck and WHT are also overplotted. Notice the clear peaks around $z \approx 0.2$. The 'toothbrush' H$\alpha$ emitter distribution peaks at a slightly higher redshift than the `Sausage' which is in line with the design of the NB filters. The ranges of higher redshift emitters possibly captured by our NB filters are marked in green. More high redshift emitters are captured in the `Toothbrush' field, given its slightly deeper data.}
\label{fig:photoz}
\end{figure}

As we do not have spectroscopy for all sources, we compute photometric redshifts. Using the comprehensive optical photometry available for these two fields (B, g, V, r, i, z, NB), relatively precise photometric redshifts (with errors $\Delta z_\mathrm{phot}/z_\mathrm{phot}<20$ per cent) can be derived. For this purpose, we performed a grid-based redshift search between $0.01$ and $1.3$ with the \textsc{eazy} code \citep{2008ApJ...686.1503B}. Full freedom has been otherwise allowed in the fitting process of the full set of PEGASE 2.0 templates \citep[described in][]{2006A&A...449..951G}, which includes a range of early to late type galaxies, with a range of stellar ages. We used magnitudes measured in $5$ arcsec apertures.
% and, to be conservative, the magnitude errors reported by \textsc{sextractor} were tripled. 
%We triple the errors since \textsc{sextractor} only takes into account the background RMS noise when calculating the error on the magnitude, without accounting for other possible sources of uncertainty such as the error in the determination of the ZP. 
The large apertures bias against the detection of high-redshift emitters, which are more compact and faint and are likely missed by our selection. Therefore, we expect our emitter population to be predominantly H$\alpha$ emitters at $z\sim0.2$. The distribution of photometric redshifts $z_\mathrm{phot}$ can be found in Fig.~\ref{fig:photoz}. A natural spread in the redshifts is expected given the uncertainties in fitting photometric redshifts, especially for star-forming galaxies, which can be relatively featureless in the continuum, e.g. they do not have strong $4000${\AA} breaks. The majority of the line emitters ($>85\%$) are correctly fit with templates that include line emission features, while galaxies at $z\sim0.8-0.9$ are fit with passive galaxy templates marked by absorption features. Therefore, not only the redshifts, but also the correct spectral type can be recovered from the template fitting.

The effect of Galactic extinction (see Figs.~\ref{fig:dust:sausage} and \ref{fig:dust:toothbrush}) is evident in the photometric redshift fitting: if we use magnitudes uncorrected for dust attenuation, the bulk of the line emitters are fitted as higher redshift passive galaxies. Hence, as explained in Section~\ref{sec:obs-reduction:red}, correcting for Milky Way dust is of the utmost importance.

Emitters are selected as H$\alpha$ emitters if the photometric redshift lies between $0.16$ and $0.23$. H$\beta$/[O{\sc iii}] explains the emission if $0.5<z_\mathrm{phot}<0.7$, [O{\sc ii}] if $1.0<z_\mathrm{phot}<1.2$ and $4000${\AA} break galaxies if $0.7<z_\mathrm{phot}<0.9$ (Figs.~\ref{fig:colcol} and \ref{fig:photoz}).

\subsubsection{H$\alpha$ emitter selection}\label{sec:hasel}
A source is selected as an H$\alpha$ emitter if it fulfils any of the criteria listed below (see Fig. \ref{fig:colcol}). We give the number of sources selected through each criterion for each field in the parentheses. Note that the spectroscopic confirmation overlaps with the other criteria in most cases, confirming the robustness of our selection.
\begin{itemize}
\item passes the colour-colour selection (equation \ref{eq:colcol}, either has clear detections in all bands B, g, r, i or an upper limit in B, i.e. the source is not detected in the B band below the detection limit in the B band, see Table~\ref{tab:limmag}) and a $z_\mathrm{phot}$ placing it at the cluster redshift ($63$ sources for the `Sausage' and $24$ sources for the `Toothbrush' field) \textit{or}
\item passes the colour-colour selection (either has clear detections in all bands B, g, r, i or significant upper limit in B), but not the $z_\mathrm{phot}$ criterion due to insecure $z_\mathrm{phot}$ (the $\chi^2$ distribution does not have a clear minimum around $z\sim0.2$, with minima of similar significance at other redshifts; we selected sources with a difference greater than $0.15$ between the primary redshift solution and the redshift marginalised over $p(z|C)=exp(-0.5\chi^2)$, or $z_\mathrm{m1}$ as denoted in \texttt{eazy}) ($42$ sources for the `Sausage' and $15$ sources for the `Toothbrush' field) \textit{or}
\item does not pass the colour-colour selection, but has a secure $z_\mathrm{phot}$ (the $\chi^2$ distribution has a clear minimum around $z\sim0.2$, without minima of similar significance at other redshifts, i.e. difference between the primary redshift solution and $z_\mathrm{m1}$ is lower than $0.15$) ($2$ sources for the `Sausage' and $11$ sources for the `Toothbrush' field) \textit{or}
\item $z_\mathrm{spec}$ confirms it is at the cluster redshift ($65$ sources for the `Sausage' and $4$ sources for the `Toothbrush' field).
\end{itemize}
The fraction of spectroscopically confirmed H$\alpha$ emitters in the `Sausage' is extremely high ($\sim52.5$ per cent, see Fig. \ref{fig:photoz}). The location of the spectroscopically confirmed sources fully validates the colour-colour selection (Fig. \ref{fig:colcol}). Based on the colour-colour selection the photometric redshift catalogue suffers from $20$ per cent mis-classifications. Most of these are sources where the photometric redshift probability distribution was roughly equal for classifying the source as H$\alpha$, H$\beta$/[O{\sc iii}] or [O{\sc ii}]. In the case of the `Sausage' field, out of $95$ sources that would be classified as H$\alpha$ by the photo-z, $37$ were targeted by spectroscopy. Out of $76$ emitters confirmed through spectroscopy, $40$ were also correctly classified by the photometric redshift method. Another $11$ sources were assigned as H$\beta$/[O{\sc iii}] or [O{\sc ii}] emitters, instead of their right type. For the rest of the sources photometric redshifts between $0.05$ and $0.35$ were assigned. In the case of the `Sausage' cluster, out of $129$ potential H$\alpha$ emitters selected through the criteria above, $5$ were removed as confirmed stars of higher redshift emitters (amounting to a contamination of less than $<4$ per cent). The emitters were also visually inspected to check for possible interlopers and obvious H$\alpha$ emitters not selected by our method. The visual inspection indicates a rate of $\sim10$ per cent possible H$\alpha$ sources not categorised as H$\alpha$ by our method (or $90$ per cent incompleteness), which is similar to \citet{2008ApJS..175..128S}.

We select a total of $124$ H$\alpha$ emitters located at the `Sausage' cluster redshift and $50$ for the `Toothbrush' FOV (Table~\ref{tab:emitters} and Figs.~\ref{fig:colcol} and \ref{fig:image}). For similar H$\alpha$ luminosities, the typical fraction for H$\alpha$ emitters at $z\sim0.2$ out of a population of emitters selected in blank fields with a NB filter is $\sim15-20$ per cent \citep[e.g.][]{2008ApJS..175..128S}. The fraction of H$\alpha$ emitters (based on Table~\ref{tab:emitters}) in the `Toothbrush' FOV resembles that of blank fields ($11$ per cent, $50$ H$\alpha$ out of $463$ emitters), while in the `Sausage' the fraction ($\sim62$ per cent, $124$ out of $201$) is significantly above field levels.

\subsection{Removing [N{\sc ii}] contamination}\label{sec:NII}

The [N{\sc ii}] forbidden line is very close in wavelength to the H$\alpha$ line ($\sim20$ {\AA} away, restframe). Since our filters are $110$ {\AA} wide, we expect to pick up emission from both the H$\alpha$ and [N{\sc ii}]. We remove the contribution to the line flux, using the relation from \citet{2012MNRAS.420.1926S}. The median contribution of [N{\sc ii}] to the H$\alpha$+[N{\sc ii}] flux is $0.32$, consistent with solar metallicity. 

\subsection{H$\alpha$ luminosity}
The H$\alpha$ luminosity $L_{\mathrm{H}\alpha}$  can be calculated from the H$\alpha$ flux $F(\mathrm{H}\alpha)$, corrected for [N{\sc ii}], as described in \S\ref{sec:NII}.  
\begin{equation}
\label{eq:L}
L_{\mathrm{H}\alpha}=4 \pi d^2_{L} F(\mathrm{H}\alpha).
\end{equation}
where $d_{L}$ is the luminosity distance ($941$ Mpc for a redshift of $0.1945$ and $1107$ Mpc for $0.2250$, respectively for the two clusters). The emitters are binned based on their luminosity and normalised by the survey volume to form a luminosity function. 

\subsection{Completeness correction}\label{sec:results:completeness}
Fainter H$\alpha$ emitters and the emitters with lower line EWs will not enter our H$\alpha$ emitter catalogue given our emitter selection criteria on limiting $\Sigma$ and $EW$. This results in incompleteness. We study the way our completeness rate varies as a function of line luminosity following the method of \citet{2012MNRAS.420.1926S}. For this, we pass sub-samples of our emitters population through our selection criteria for H$\alpha$ emitters described in Section~\ref{sec:hasel}. 

We select a sample of sources, consistent with being non-emitters, but which pass our colour-colour criteria as being located at the cluster redshift. We add fake H$\alpha$ emission lines to the flux of these galaxies and fold them through our $EW$ and $\Sigma$ emitter selection criteria (as shown in \S\ref{sec:obs-reduction:NBselection}) and study the recovery fraction. We perform this study independently for eight areas, given the variation of the dust extinction across the FOV \citep[as noted by][see also Fig.~\ref{fig:dust:sausage}]{2014MNRAS.438.1377S}. The variable dust extinction has a non-trivial effect on the recovery rate of H$\alpha$ emitters, given the different way dust extinction affects the blue side and the red side of the spectrum, affecting the perceived colours of the emitters.

\begin{figure*}
\begin{center}
\includegraphics[trim=0cm 0cm 0cm 0cm, width=0.495\textwidth]{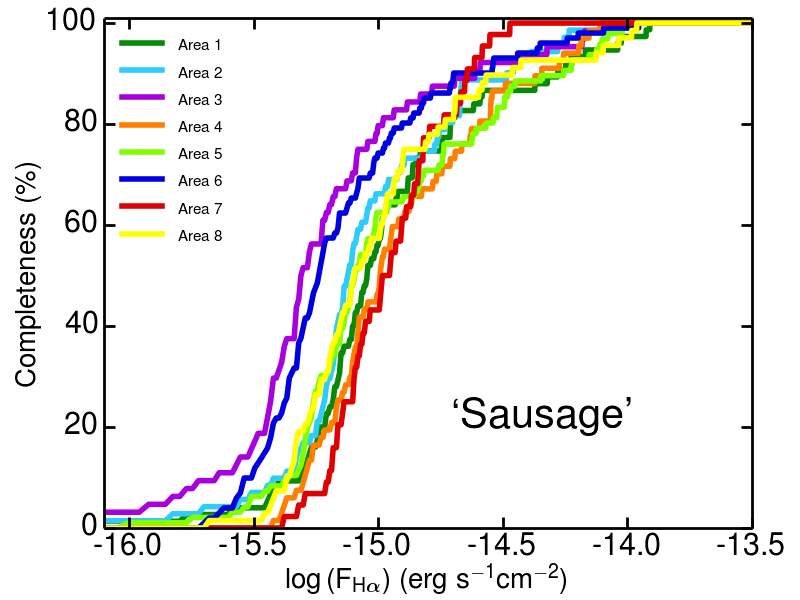}
\includegraphics[trim=0cm 0cm 0cm 0cm, width=0.495\textwidth]{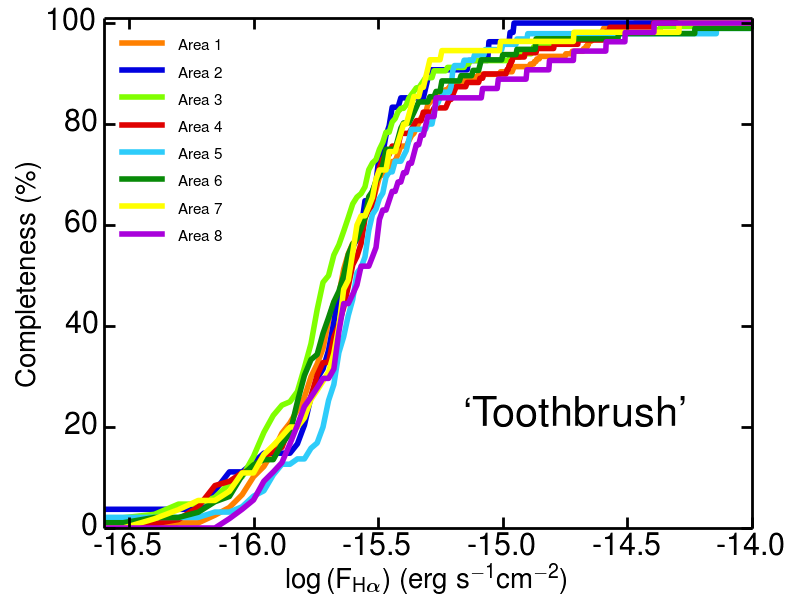}
\end{center}
\caption{Survey completeness for the `Sausage' (left) and `Toothbrush' (right) fields as a function of H$\alpha$ flux, performed separately for eight sub-areas within the FOV, to account for the varying dust extinction (see Figs.~\ref{fig:dust:sausage} and \ref{fig:dust:toothbrush}). The areas are colour-coded according the dust extinction, from low dust extinction in indigo, though blue, green yellow, orange and red for the highest extinction. Note the correlation between completeness and amount of extinction: areas 3 and 6 in the `Sausage' field have the highest level of completeness, being the least affected by dust.}
\label{fig:completeness}
\end{figure*}

The recovery rate of H$\alpha$ emitters as function of their flux, for the eight areas of the FOV can be found in Fig.~\ref{fig:completeness}.

\subsection{Volume and filter profile corrections}
Assuming a top-hat (TH) shape with FWHM of $110$ {\AA} for the NB filter transmission profiles and given the FOV coverages (see Fig.~\ref{fig:image}), we are surveying a total comoving volume of $3.435\times10^3$~Mpc$^3$ for the `Sausage' and $4.625\times10^3$~Mpc$^3$ for the `Toothbrush'. Since our filters are not perfect THs (Fig.~\ref{fig:transmittance}), we correct the volumes for the possible sources which might be missed in the wings of the filter as shown in \citet{2009MNRAS.398...75S, 2012MNRAS.420.1926S}. 

\subsection{Survey limits}\label{sec:results:limits}
We probe luminosities down to a $10$ per cent completeness limit (see Fig.~\ref{fig:completeness}). This is equivalent to an average limiting H$\alpha$ luminosity of $10^{40.64}$ erg s$^{-1}$ for the `Sausage' field and $10^{40.14}$ erg s$^{-1}$ for the `Toothbrush' field. The actual limiting magnitude will vary across the FOV much as the completeness varies, due to the varying dust extinction. Assuming all the H$\alpha$ luminosity comes from star formation, we can use the \cite{1998ARA&A..36..189K} relation, with a \citet{2003PASP..115..763C} initial mass function (IMF) to obtain the limiting star formation rate (SFR) of our surveys (see equation \ref{eq:SFR} in Section \ref{sec:SFRD}). The `Sausage' field average limiting SFR is $0.17$ M$_{\sun}$ yr$^{-1}$, while for the 'toothbrush' we reach down to an average of $0.06$ M$_{\sun}$ yr$^{-1}$. 

\subsection{H$\alpha$ luminosity function and star formation rate}
We fit LFs to different regions within the two clusters (e.g. relic areas that are aligned with the merger axis, the sides of the cluster perpendicular to the axis where there is no radio emission, weak lensing area where most of the mass is contained etc.). For this purpose, we bin the data based on luminosity over each area. We bin the data using a range of bins. We define the bins by varying the minimum luminosity and bin width ($100$ random choices with uniform distribution). The errors on the $\log \phi$ values are Poissonian. By resampling the LF in different ways we can obtain a more robust determination of the fit parameters, which are not dominated by a particular choice of binning. To compare with other studies, we use the popular parametrisation of the LF defined by \citet{1976ApJ...203..297S}:
\begin{equation}
\label{eq:schechter}
\phi(L) \mathrm{d} L = \phi^*\left(\frac{L}{L^*}\right)^{\alpha} e^{-(L/L^*)} \mathrm{d} \left(\frac{L}{L^*}\right),
\end{equation}
where $L^*$ is the characteristic luminosity of the emitters where the power-law cuts off. $\phi^*$ is the density of H$\alpha$ emitters and provides the normalisation. $\alpha$ is the faint-end slope of the LF, which we fixed to $-1.2$ \citep{2014MNRAS.438.1377S}. 

\begin{figure*}
\begin{center}
\includegraphics[width=0.97\textwidth]{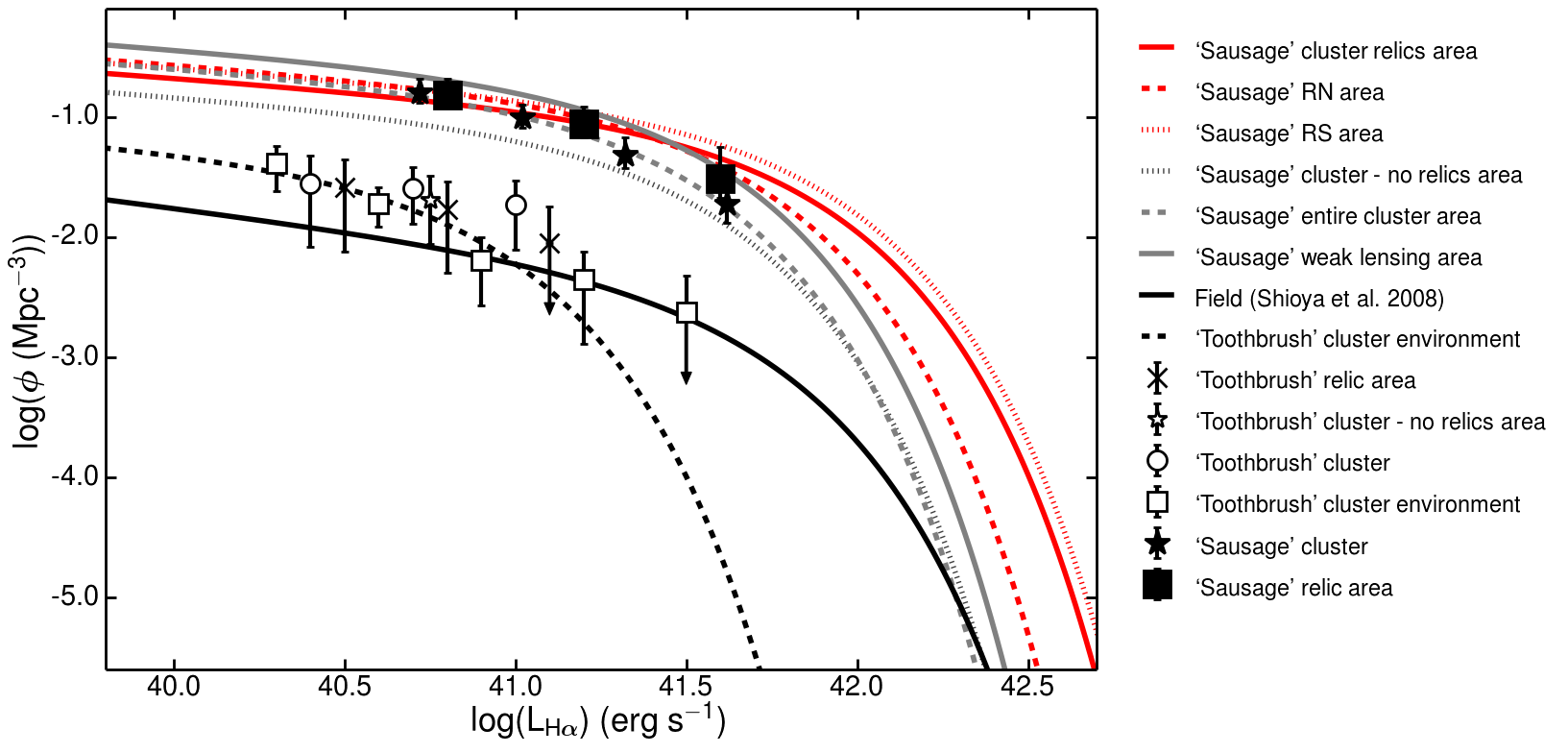}
\end{center}
\caption{Luminosity function for the two clusters, obtained by fixing the faint end slope $\alpha$ to $-1.2$ \citep[as derived in][]{2014MNRAS.438.1377S} and varying the normalisation $\log \phi^*$ and the characteristic luminosity $\log L^*$. The regions are defined according to Fig.~\ref{fig:image}. Downward pointing arrows represent upper limits equivalent to $1$ source per volume element, per bin. The fit for a $z=0.24$ blank field from \citet{2008ApJS..175..128S} is oveplotted. Note the difference in normalisation between the two clusters: the density of H$\alpha$ emitters in the `Sausage' is a factor of $10$ higher than in the `Toothbrush'. The typical luminosity $L^*$ of `Toothbrush' emitters is lower than the blank field.}
\label{fig:LF}
\end{figure*}

We produce a 2D distribution of number of realisations as function of LF parameters (i.e. how many of the randomly generated binnings were best fitted with a particular combination of $\phi^*$ and $L^*$). The reported $\log \phi^*$ and $\log L^*$ are determined as the mean of this distribution obtained by resampling the LF. The errors are $1\sigma$ standard deviations away from the mean. The results of the fit for different regions within and around the clusters (as defined in Fig. \ref{fig:image}) can be found in Table~\ref{tab:LF} and Fig.~\ref{fig:LF}. There is a striking difference between the normalisation of the two cluster LFs, as discussed in more detail in Section~\ref{sec:discussion}. The `Toothbrush' LF is similar to a blank field, while the `Sausage' $\phi^*$ is about a factor of $10$ larger than that (Fig.~\ref{fig:LF}).

We can use the Kennicutt (1998) conversion from H$\alpha$ luminosity to star formation activity, using a Chabrier IMF \citet{2003PASP..115..763C} IMF:
\begin{equation}
\label{eq:SFR}
SFR(\mathrm{M_\odot} \mathrm{yr^{-1}}) = 4.4 \times 10^{-42} L_{H\alpha} \mathrm{(erg s^{-1})}.
\end{equation}

\subsection{Stellar masses}\label{sec:masses}

Galaxy spectral energy distributions were generated with the \citet{2003MNRAS.344.1000B} software package. We used stellar synthesis models from \citet{2007ASPC..374..303B}, a Chabrier IMF with exponentially declining star formation histories and a range of metallicities. These models were fitted to the full broad-band data (BgVrIz) to obtain stellar masses for the H$\alpha$ galaxies, following the method presented in \citet{2011MNRAS.411..675S,2014MNRAS.437.3516S} (for further details, see Sobral et al in prep). A histogram of the masses of H$\alpha$ emitters in the `Sausage' and `Toothbrush' field is shown in Fig.~\ref{fig:histmass_norm}. The values are normalised by the volume of each survey. The masses of the `Sausage' H$\alpha$ emitters are on average $\sim10^{9.8}$\,M$_{\odot}$. On average, the `Toothbrush' H$\alpha$ emitters are $\sim3$ times less massive than those in the `Sausage' ($\sim10^{9.3}$\,M$_{\odot}$). Note that this is driven by the deeper `Toothbrush' data: the faintest H$\alpha$ flux detected in the `Sausage' survey is $3.4\times10^{-16}$ erg s$^{-1}$ cm$^{-2}$, while in the `Toothbrush' we probe down to $4.2\times10^{-17}$ erg s$^{-1}$ cm$^{-2}$, a factor of $\sim8$ deeper. Note that despite the similar volumes probed by the two surveys, there are significantly fewer massive H$\alpha$ emitters in `Toothbrush' compared to the `Sausage'.

In Fig.~\ref{fig:sfrmass}, we plot the SFR versus the mass of the cluster H$\alpha$ emitters, on top of the results for the blank field obtained from COSMOS \citep{2008ApJS..175..128S}. We compute SFR for individual galaxies using the dust correction based on stellar mass \citep{2010MNRAS.409..421G}. While our cluster H$\alpha$ emitters fall on the SFR-mass relationship as defined from blank fields, it is important to note that the COSMOS data was obtained over a volume $\sim10$ times higher than the volume probed by our `Sausage' and `Toothbrush' surveys. We are detecting high numbers of very high-mass, highly-star forming galaxies which are relatively rare in the field.

\subsection{The star formation rate density}\label{sec:SFRD}
Given a LF and using the conversion from H$\alpha$ luminosity and SFR, we can also calculate the star formation rate density $\rho_{SFR}$ within that particular volume. The luminosity density is obtained by integrating the LF:
\begin{equation}
\rho_L = \int_0^{\infty} \phi(L)L \mathrm{d}L =  \Gamma(\alpha+2) \phi^* L^*,
\end{equation}
where $\Gamma(n)=(n-1)!$ is the gamma function.
$\rho_\mathrm{SFR}$ is then: 
\begin{equation}
\label{eq:sfrd}
\rho_\mathrm{SFR} =  \Gamma(\alpha+2) 10^{\phi^*} 10^{L^*} 10^{0.4} (1-f_\mathrm{AGN})
\end{equation}
where we are assuming that part of the H$\alpha$ emitters are powered by active galactic nuclei, rather than star formation. From the spectroscopic observations of the `Sausage' H$\alpha$ emitters, emission lines ratios (comparison of [O{\sc iii}]/H$\beta$ versus [N{\sc ii}]/H$\alpha$ \citep[BPT diagram;][]{1981PASP...93....5B} enable the separation between star-forming galaxies and AGN (Sobral et al. in prep). This analysis indicates a fraction of $10$ per cent pure AGN-powered H$\alpha$ emitters. We thus assume that a fraction $f_\mathrm{AGN}$ of the emitters is powered by AGN activity, chosen to be $10$ per cent according to Sobral et al (in prep) and in line with \citet{2010MNRAS.409..421G} and \citet{2013MNRAS.428.1128S}. We also correct the luminosity for the dust extinction intrinsic to the line emitter. A conservative value of $1.0$ magnitude has been shown to be appropriate by various authors \citep[e.g.][]{2012MNRAS.420.1926S,2013MNRAS.434.3218I} and has been widely assumed in the literature \cite[e.g.][]{2003ApJ...586L.115F,  2007ApJ...657..738L, 2008MNRAS.388.1473G, 2009MNRAS.398...75S}. The error on the $\rho_\mathrm{SFR}$ is $\Delta\rho_{SFR}$:
\begin{equation}
\label{eq:sfrderr}
\Delta\rho_\mathrm{SFR} = \rho_\mathrm{SFR} \ln(10) \sqrt{(\Delta\phi^*)^2+(\Delta L^*)^2}
\end{equation}
The comparison of $\rho_\mathrm{SFR}$ in and around our clusters and the field is illustrated in Fig.~\ref{fig:SFRD}. We also overplot the value obtained by \citet{2008ApJS..175..128S} for the COSMOS field. For comparison, we show the parametrisation derived by \citet{2013MNRAS.428.1128S} using a method similar to ours from a set of consistent H$\alpha$ NB observations at four redshift slices ($2.23$, $1.47$, $0.84$ and $0.40$). The dependence of the $\rho_\mathrm{SFR}$ on redshift for the blank field is:
\begin{equation}
\log_{10} \rho_\mathrm{SFR} = -2.1 (1+z)^{-1}.
\end{equation}
We find that $\rho_\mathrm{SFR}$ for the Sausage cluster is significantly enhanced compared to the field. See Section~\ref{sec:discussion} for a more detailed discussion.

\section{Discussion}\label{sec:discussion}
The fraction of star-forming galaxies drops steeply from field environments towards the cores of massive, hot, relaxed clusters \citep[e.g.][]{2003ApJ...584..210G}. To study the effect of cluster mergers and shocks on the star forming properties of cluster galaxies, we performed observations of two disturbed, $z\sim0.2$ clusters hosting radio relics, tracers of ICM shock waves of Mach number $3-4$. We look at the trends in the normalisation and specific luminosity of the H$\alpha$ luminosity function for the clusters and the masses, star-formation rate and star-formation rate density for the cluster star-forming galaxies. We discuss possible scenarios in which we expect differences between the H$\alpha$ population of the two clusters. 

\begin{table*}
\begin{center}
\caption{Parameters of the luminosity functions fitted with a Schechter function (see equation~\ref{eq:schechter}). The faint-end slope was fixed to a value of $\alpha=-1.2$, which was obtained by \citet{2014MNRAS.438.1377S} by combining data for the entire FOV of the two clusters. The fits were possible only for some areas within the cluster, as the number statistics were not in all cases high enough to allow the fitting of the independent Schechter parameters. For each area, the data was binned in range of different ways (by varying the luminosity of the first bin and the bin width) and fit independently to obtain an average, characteristic fit that best describes the LF shape. The average and standard deviation over these independent fits is reported in the table. The blank field fit for the COSMOS field by \citet{2008ApJS..175..128S} is also reported. We also list $\rho_\mathrm{SFR}$ corresponding to each fit. The number of H$\alpha$ emitters employed in each fit is given in the last column.}
\begin{tabular}{l l c c c c c}
\hline
\hline
Field                       & $\alpha$ & $\log \phi^*$ & $\log L^*$ & SFRD & Number of H$\alpha$ & $z_\mathrm{spec}$ confirmed H$\alpha$\\ 
							 &			& Mpc$^{-3}$ &  erg s$^{-1}$ &  $M_{\odot}$ yr$^{-1}$ Mpc$^{-3}$      & & \\\hline  \vspace{3pt}
`Sausage' field & & & &  & \\ \hline \vspace{3pt} 
Cluster relics area & $-1.2$    &  $-1.37\pm0.04$   & $41.69\pm0.09$  & $0.22\pm0.05$   & $32$  & $15$ \\ \vspace{3pt}
RN area  			& $-1.2$    &  $-1.22\pm0.04$   & $41.51\pm0.06$  &  $0.21\pm0.04$  & $20$  & $7$ \\ \vspace{3pt}
RS area             & $-1.2$    &  $-1.29\pm0.04$   & $41.72\pm0.19$  &  $0.29\pm0147$  & $12$  & $8$ \\ \vspace{3pt}
Weak lensing area   & $-1.2$    &  $-1.07\pm0.04$   & $41.40\pm0.05$   & $0.23\pm0.03$  & $26$  & $19$ \\ \vspace{3pt}
Entire cluster area & $-1.2$    &  $-1.21\pm0.02$   & $41.33\pm0.02$   & $0.14\pm0.01$  & $59$  & $30$ \\ \vspace{3pt}
Cluster - no relics area & $-1.2$    &  $-1.46\pm0.03$   & $41.38\pm0.06$ & $0.09\pm0.01$ & $14$ & $5$  \\ \hline\hline \vspace{5pt} 
`Toothbrush' field & & & &  & \\ \hline \vspace{3pt} 
Cluster environment & $-1.2$  &  $-1.76\pm0.04$   & $40.75\pm0.01$   & $0.01\pm0.001$  & $25$  & $4$ \\ \hline\hline \vspace{5pt}
\citet{2008ApJS..175..128S} & $-1.35^{+0.11}_{-0.13}$    &  $-2.65^{+0.27}_{-0.38}$   & $41.57^{+0.38}_{-0.23}$ & $0.018^{+0.007}_{-0.004}$ & &  \\
\hline
\end{tabular}
\label{tab:LF}
\end{center}
\end{table*}

\subsection{H$\alpha$  luminosity function for the clusters}\label{sec:LF:sausage}

\subsubsection{`Sausage' H$\alpha$ luminosity function}
In the case of the `Sausage', we build independent H$\alpha$ LFs for the areas around the relics, along the merger axis of the cluster, and also away from the relics, perpendicular to merger axis (see Fig.~\ref{fig:image}). We study the LF for the entire cluster, by encompassing it with a circular aperture of $\sim1.85$ Mpc radius. This radius was chosen as function of the X-ray temperature peaks \citet{2013MNRAS.429.2617O} and the positions of the radio relics, which should be located at the cluster outskirts. A weak lensing analysis of the cluster \citep{Jee}, indicates a value of $r_{200}=2.63$ Mpc for the radial extent, assuming the cluster is fitted with a single Navarro-Frenk-White dark matter halo \citep{1997ApJ...490..493N}. Therefore, our chosen radius for the cluster is smaller than $r_{200}$. We also build LFs for the area where most of the weak lensing mass in contained, as per \citet{Jee}. We smooth the luminosity density map with a Gaussian kernel of $125$ arcsec. We consider the weak lensing area to be contained within the contour defined by a $0.4$ value of the peak in the smoothed map. This corresponds to a mass of about $1\times10^{15}$ $M_{\odot}$ (half the mass of the entire cluster) that is enclosed within this contour.

\begin{figure}
\begin{center}
\includegraphics[width=0.47\textwidth]{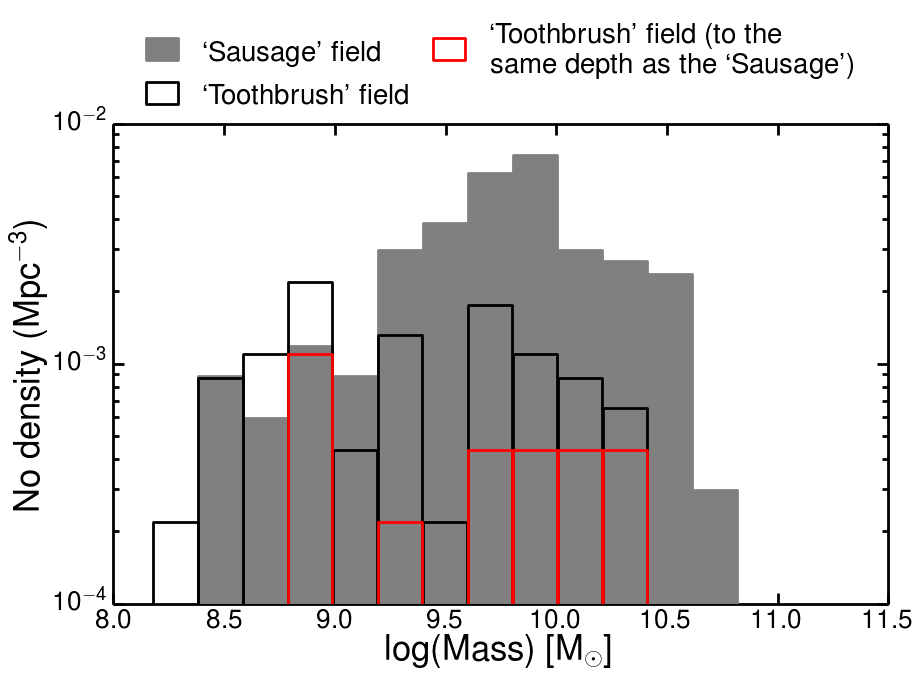}
\end{center}
\caption{Mass distribution of H$\alpha$ emitters in the fields of the two clusters, normalised to the volume of each respective survey. Note than even though the `Toothbrush' data probes to H$\alpha$ fluxes fainter by a factor of $8$ compared to the `Sausage', the `Toothbrush' cluster is almost devoid of massive star-forming galaxies.}
\label{fig:histmass_norm}
\end{figure}

The normalisation of the H$\alpha$ LF for all areas within the `Sausage' cluster is higher by a factor $>15$ compared to the fit for a blank field at $z=0.24$ by \citet{2008ApJS..175..128S}. The field LF parameter fits are expected to be only moderately affected by cosmic variance given the volume probed ($3.1\times10^{4}$ Mpc$^3$) and the small correlation length of $r_0=1.8\pm0.2$ Mpc of the observed H$\alpha$ emitters at $z=0.24$, roughly corresponding to emitters residing in typical dark matter halos of $10^{11}M_\odot$ masses \citep{2010MNRAS.404.1551S}. There is evidence that the number of emitters within the weak lensing area, which traces the direction of the merger, is enhanced compared to other areas within the cluster. The LF normalisation $\log \phi^*$ in the weak lensing area is $-1.07\pm0.04$ compared to an average of $-1.37\pm0.04$ near the relics (difference of $5.0\sigma$) and $-1.46\pm0.03$ in the region away from the relics ($7.8\sigma$), perpendicular to the merger axis. This fully confirms the results from \citet{2014MNRAS.438.1377S}, where, even if a conservative, low fraction of H$\alpha$ emitters was considered, an enhancement in the $\log \phi^*=-1.77$ around the relics area was observed by comparison to field galaxies ($\log \phi^*=-2.65$). 

These very high numbers of H$\alpha$ galaxies within the densest parts of the `Sausage' cluster are surprising, given the number of studies showing that the fraction of star-forming galaxies drops steeply towards the cluster core. Even in the case of medium redshift clusters such as the $z=0.81$ RXJ1716.4+6708 cluster, \citet{2010MNRAS.403.1611K} have found that the fraction of H$\alpha$ emitters drops steeply towards the regions of highest density, down to values of just $10\%$ of the total population. The authors use a similar method to ours: a narrow-band H$\alpha$ selection technique to identify star forming galaxies. A different trend is found at higher redshifts, in dynamically young clusters. For example, \citet{2011MNRAS.415.2670H} have found that at $z\sim1.5$, the centre of a dynamically young cluster is still full of star-forming galaxies and AGN. Therefore, the `Sausage' cluster presents features very similar to a young cluster at the time where the progenitors of the most massive clusters have started forming.

Note that we calculate the volume by taking the entire redshift slice captured by the NB filter. The volume occupied by the cluster which, given its projected dimension, is expected to be $<3$ Mpc (real-space) across in the redshift direction. If we were to assume the cluster occupies a cylindrical volume, with a circular shape projected onto the sky, the cluster volume would be a maximum of $\sim32$~Mpc$^3$ (real-space volume). We subtract the typical level of field emitters \citep[$\log \phi^*=-2.65$, as per][]{2008ApJS..175..128S} and consider all additional H$\alpha$ emitters projected on the cluster area to be cluster members. Note that this method assumes that there are field galaxies mixed within the cluster volume, but this is not expected to make a big difference in the final numbers. This exercise leads to an enhancement by $\sim1.5$ dex (a factor of $30$) compared to the $\log \phi^*$ calculated in Table \ref{tab:LF} ($0.10$ versus $-1.27$). If the background level of field H$\alpha$ galaxies is not subtracted $\log \phi^*$ is $0.12$. The same enhancement would be seen in all areas if the volume was adjusted.

Even more so, there are a high number of spectroscopically confirmed H$\alpha$ cluster members close to the relics and along the merger axis. Within the entire cluster encompassed by the large circle there are $30$ confirmed cluster members, compared to $59$ sources in total. This means that even if we build an H$\alpha$ LF using the spectroscopically confirmed members and a cluster volume of $\sim32$~Mpc$^3$, $\log \phi^*$ is $-0.18$, significantly above the field where $\log \phi^*=-2.65$. The difference lies at $>8\sigma$ significance.

\begin{figure}
\begin{center}
\includegraphics[width=0.47\textwidth]{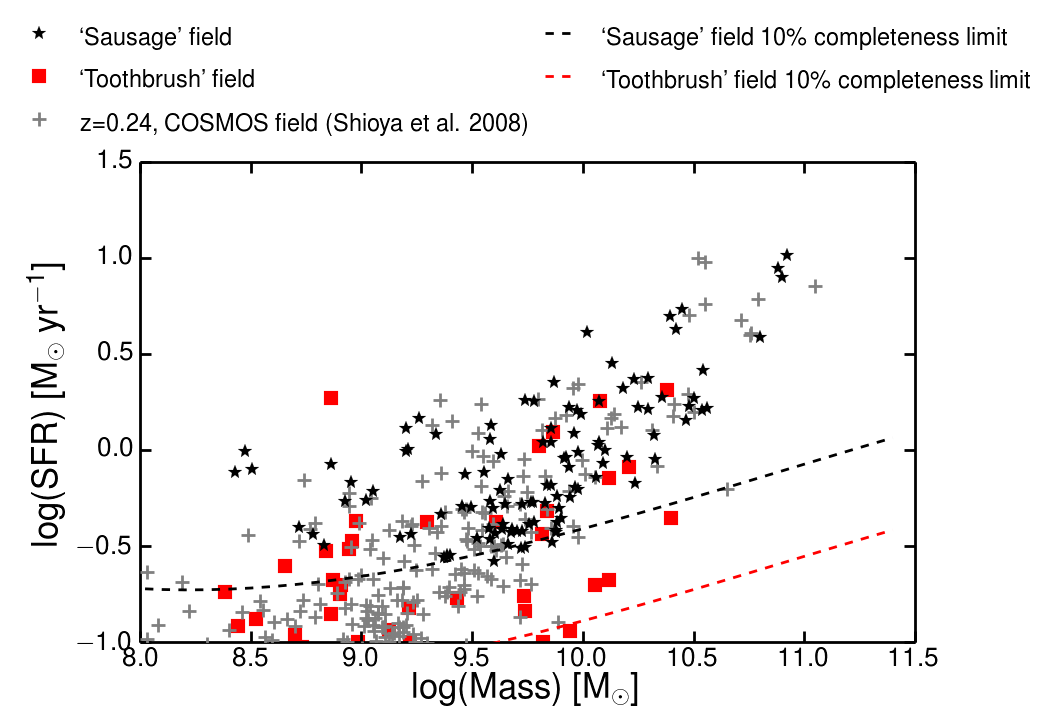}
\end{center}
\caption{Star-formation rate as function of stellar mass for the two clusters and the blank field \citep{2008ApJS..175..128S}, using a Chabrier IMF. The \citet{2008ApJS..175..128S} results for the COSMOS field were corrected for the different aperture sizes. The average $10\%$ completeness line for the `Sausage' and `Toothbrush' surveys is shown in the dashed lines. The SFR-mass relation holds for both the blank field and the two clusters. Note the overdensity of high-mass, high-SFR galaxies within the `Sausage' cluster, which surveys a volume $\sim10$ times smaller than the COSMOS field. This point is illustrated further in Figs.~\ref{fig:sausdiv} and \ref{fig:toothdiv}.}
\label{fig:sfrmass}
\end{figure}

We also detect a mild boost in the characteristic luminosity $\log L^*$ of the cluster galaxies located in proximity of the relics compared to the field ($41.69\pm0.09$ versus $41.57^{+0.38}_{-0.23}$). By contrast, the population located away from the shocks seems to have lower luminosities on average, pointing towards a quenching of luminous H$\alpha$ emitters in that direction. The difference between the H$\alpha$ emitters located around the relics and away from the relics is at the level of at least $3\sigma$. H$\alpha$ luminosity scales with SFR, which means the `Sausage' relic emitters are slightly more star-forming than field galaxies at the same redshift and cluster galaxies located away from the shock fronts. \citet{1998ApJ...504L..75B} found, from comparing a sample of field star-forming galaxies with galaxies in X-ray luminous, $0.18<z<0.55$ clusters, that the dense cluster environment suppresses the star-formation rate of galaxies. This is consistent with the galaxies located away from the shock fronts, which have average SFR $1$ $M_\odot$ yr$^{-1}$, compared to $2$ $M_\odot$ yr$^{-1}$ for the H$\alpha$ emitters located close to the radio relics.

\begin{figure}
\begin{center}
\includegraphics[width=0.470\textwidth]{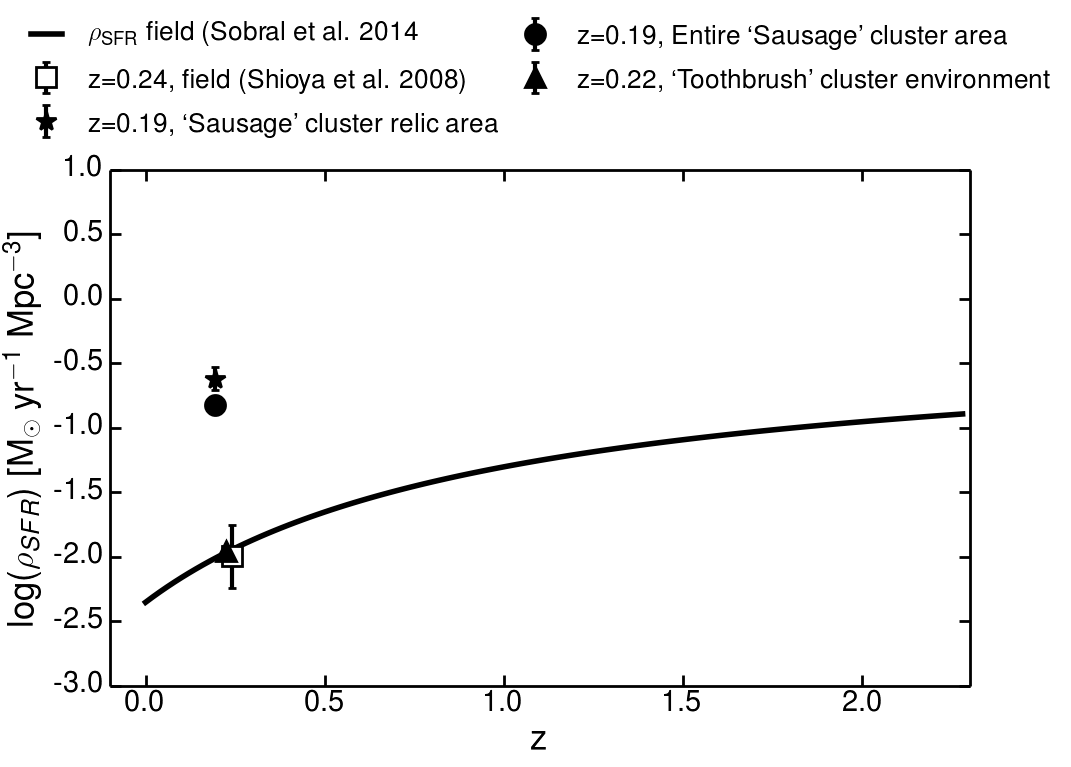}
\end{center}
\caption{$\rho_\mathrm{SFR}$ for the `Sausage' and `Toothbrush' fields plotted over the H$\alpha$ parametrisation of the $\rho_\mathrm{SFR}$ history from \citet{2013MNRAS.428.1128S}, corrected for a Chabrier IMF. The entire volume spanned by the NB filter is considered. Note that $\rho_\mathrm{SFR}$ for the `Sausage' cluster relic area is at the level of $z\sim2.3$ field galaxies. By contrast the `Toothbrush' cluster environment is at the level of $z\sim0.2$ field galaxies. The `Toothbrush' behaves more like one would expect for a cluster, where the fraction of H$\alpha$ galaxies steeply drops towards the cluster core. Error bars are plotted, but sometimes they are smaller than the symbol.}
\label{fig:SFRD}
\end{figure}

Comparison with other clusters is challenging, because there are only a few studies that use a uniform NB, H$\alpha$ selection method. \citet{2004ApJ...601..805U} derive H$\alpha$ luminosity functions for the merging cluster Abell 521, assuming that all the H$\alpha$ emitters captured by their filter are encompassed in a spherical volume of radius $2$ Mpc (with a volume of $33.5$ Mpc$^4$). The volume is therefore comparable to our method. The LF normalisation obtained by \citet{2004ApJ...601..805U} is $-0.25\pm0.20$ compared to our estimation of $0.12$. Note that the authors do not subtract a background level of field galaxies, which would make the difference between the two results more significant. The typical log10 luminosity of Abell 521 is $41.73\pm0.17$, which is consistent with the values obtained for the `Sausage' cluster relic areas. Note that the luminosity distance depends on the chosen cosmology and affects the derived H$\alpha$ luminosity, when converting from H$\alpha$ flux. To correct for this, we correct for the different cosmology used by the other authors. 

\citet{2002A&A...384..383I} derived LFs for the local cluster Abell 1367 and Coma, which have also have a lower $\log \phi^*$ of $0.06^{+0.14}_{-0.12}$ and $-0.07^{+0.03}_{-0.02}$, estimated in the same way as \citet{2004ApJ...601..805U}. The typical luminosities are $\log L^*=41.00^{+0.07}_{-0.09}$ and $\log L^*=40.97^{+0.01}_{-0.02}$, which are much lower than the overall cluster and sub-areas ($\log L^*=41.33-41.72$), at the level of $5.0-8.0\sigma$.  

Abell 1367 is a relatively cold, but dynamically active cluster with temperature averages for its two subclusters of $4.2\pm0.3$ keV and $3.2\pm0.01$ keV \citep[using Advanced Satellite for Cosmology and Astrophysics data][]{1998ApJ...500..138D}. Coma is a dynamically more evolved cluster with a low spiral fraction of $\sim13$ per cent \citep{1977ApJ...218L..93B} and an average temperature of $7.9\pm0.03$ keV \citep{1980HiA.....5..735M}. By contrast, in the `Sausage' cluster the numerous emitters along the merger axis are located within areas of extremely hot ICM. As shown in \citet{2013MNRAS.429.2617O}, the cluster temperatures along the merger axis are higher than $9$ keV, reaching $13$ keV. The enhanced numbers of SF galaxies seems to indicate that there is an inversion of the environmental trends in the `Sausage' cluster. Typically, the interaction with the hot ICM should quench the SF by removing gas from the galaxies, while the `Sausage', a dynamically active cluster, has a very high fraction of star-forming galaxies.

\begin{figure}
\begin{center}
\includegraphics[width=0.470\textwidth]{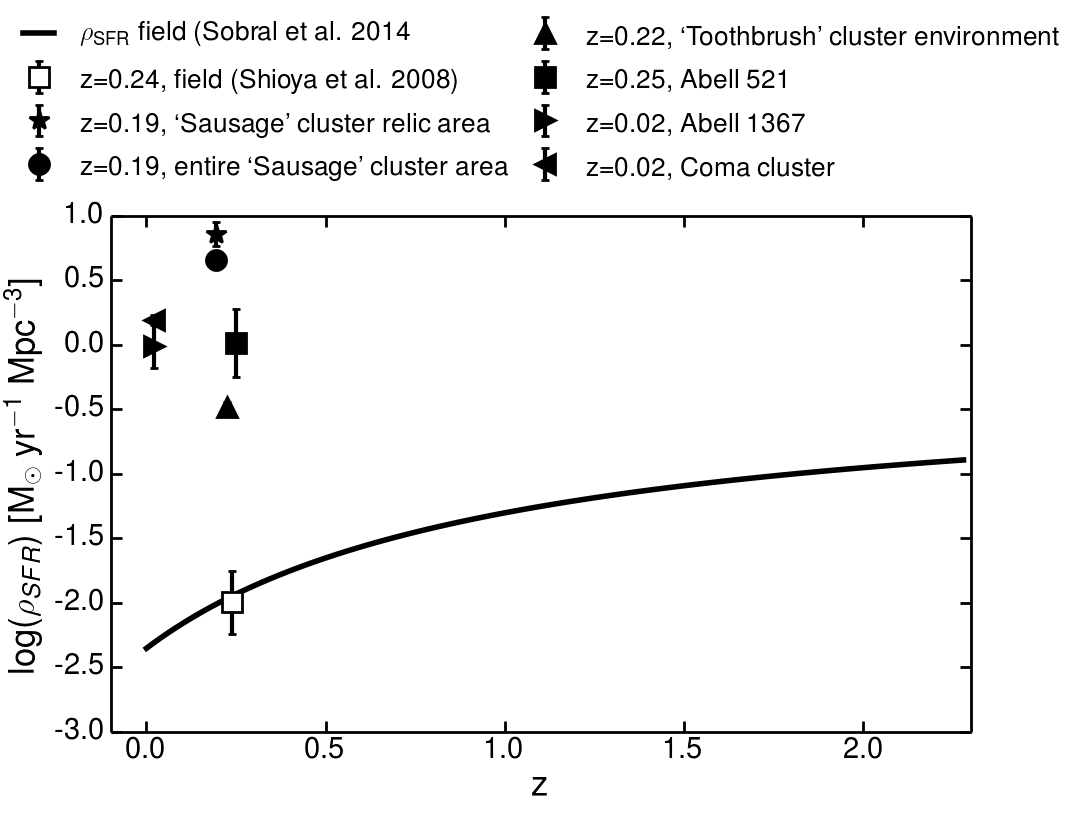}
\end{center}
\caption{$\rho_\mathrm{SFR}$ for the `Sausage' and `Toothbrush' clusters, plotted over the parametrisation of the $\rho_\mathrm{SFR}$ history from \citet{2013MNRAS.428.1128S} (corrected for a Chabrier IMF) together with Abell 521, 1367 and the Coma cluster. We correct for the cluster volume. }
\label{fig:SFRD:clust}
\end{figure}

\subsubsection{`Toothbrush' H$\alpha$ luminosity function}
In the case of the `Toothbrush', because of a lack of emitters within the cluster potential, we could not separate the cluster in multiple areas. There are no emitters in the region close to the northern relic and a few ($9$) close to the southern relic, with $3$ emitters towards the right side of the cluster, away from the relic areas. Compared to the blank field around the cluster there is an under-density of emitters within the cluster. The few H$\alpha$ emitters within the cluster region, between and around the relics are clumped together close to the southern relic. To gain enough statistics, we produce an H$\alpha$ LF only for the cluster environment encompassed by a large circular aperture of $\sim2.2$ Mpc radius, chosen to reflect the locations of the X-ray peak and the radio relics. A very preliminary weak lensing analysis of the cluster (James Jee, private communication), coupled with the X-ray temperature of this cluster \citep{2013MNRAS.433..812O}, indicates the `Toothbrush' cluster is similarly massive to the `Sausage' cluster, suggesting a similar or slightly larger radial extent of $r_{200}>2.6$ Mpc. The four spectroscopically confirmed H$\alpha$ emitters are located towards the south-west of the cluster. 

The number density of emitters is typical of what is measured for the COSMOS field \citep{2008ApJS..175..128S}. Over comparable volumes and down to the same H$\alpha$ flux, the `Toothbrush' field of view has only about $\sim12$ per cent of the number of H$\alpha$ galaxies hosted by the `Sausage'. Within the cluster volumes, down to the same H$\alpha$ flux, the `Toothbrush' cluster contains only $9$ emitters compared to $59$ in the `Sausage'. In the case of the `Toothbrush', there is evidence for a suppression of H$\alpha$ emission at high luminosities, as indicated by the lower value of the specific luminosity $\log L^*$ compared to the field ($40.75$ vs $41.57$). Nevertheless, because of low number statistics, the error in the `Toothbrush' $\log L^*$ measurement is high. The suppression of bright H$\alpha$ emission is consistent with the dense, hot environment in which galaxies are residing. X-ray observations of the `Toothbrush' cluster from \citet{2013MNRAS.433..812O} indicate that the cluster, while highly disturbed in its temperature structure, has temperatures towards the southern part of the cluster ranging between $6-8$ keV.  

Following the procedure described in Section \ref{sec:LF:sausage}, we subtract the background level of field emitters and assume all other H$\alpha$ emitters located in projection in the cluster area are actually cluster members and we adjust the volume as we did in Section \ref{sec:LF:sausage} (the cluster volume is $\sim46$~Mpc$^3$), pushing the LF normalisation to a value of $\log L^* = -0.41$. 

\begin{figure}
\begin{center}
\includegraphics[width=0.470\textwidth]{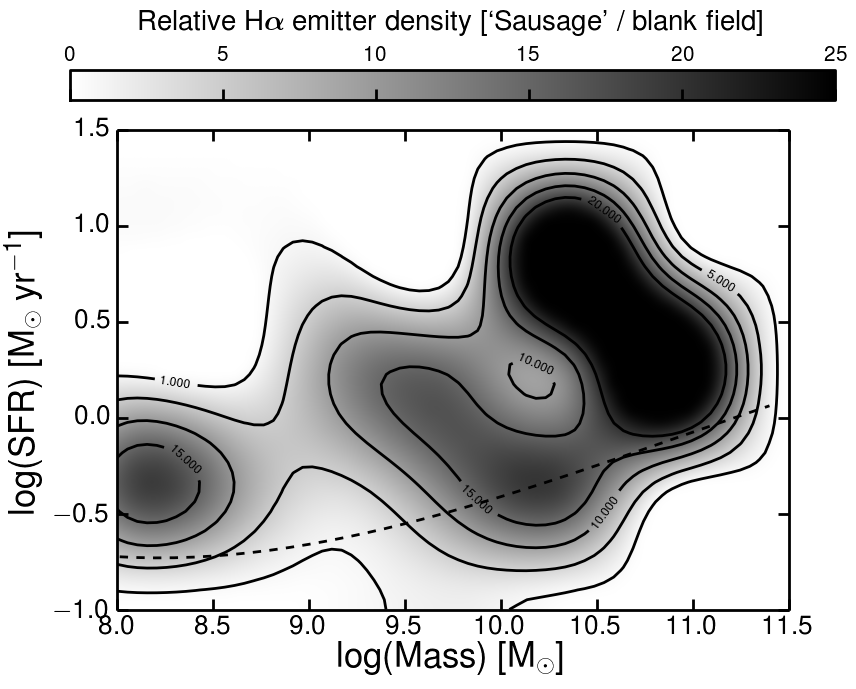}
\end{center}
\caption{Relative density of H$\alpha$ emitters of the `Sausage' field compared to a blank field from \citet{2008ApJS..175..128S}, plotted in the stellar mass - star-formation rate plane. The $10\%$ completeness in the SFR measurements is plotted in the dashed line. The `Sausage' field has a higher density of H$\alpha$ emitters at all SFRs and stellar masses, as compared to a blank field. The `Sausage' cluster contains up to $15-20$ times more highly star-forming, high-mass galaxies than a blank field.}
\label{fig:sausdiv}
\end{figure}

\subsection{Star formation rate density for the two clusters}

The $\rho_\mathrm{SFR}$ values for the `Sausage' relic area and the entire cluster area are significantly above what is expected for galaxies at redshift $0.2$ (Fig.~\ref{fig:SFRD}). Note that these values are obtained if one uses the entire redshift span of the filter, hence using co-moving volumes. The emitters in the cluster and the relic area behave on average like typical blank field galaxies at $z\sim2.3$. 

If we correct for the limited real-space volume the cluster is expected to occupy (as shown in Section~\ref{sec:LF:sausage}), $\rho_\mathrm{SFR}$ for the cluster is $6.9$ $M_{\odot}$ yr$^{-1}$ Mpc$^{-3}$, more than $15$ times the level of typical galaxies located at the peak of the SFR history (Fig.~\ref{fig:SFRD:clust}). Note that here we calculate the SFR for the cluster volume using real-space volumes, while the volume for the blank fields are calculated using co-moving volumes.  If we use only the spectroscopically confirmed galaxies, $\rho_\mathrm{SFR}$ is at least at the level $7$ times the peak of the SFR. Note, however, that the volume is small and the stellar mass density added is therefore very little, equivalent to adding until $z~0.2$ just on galaxy with a stellar mass of about $4.5\times10^{10}M_\odot$ (see also Section \ref{sec:discussion:interp}).

By contrast, the lower LF normalisation and specific H$\alpha$ luminosity in the `Toothbrush' cluster drive $\rho_\mathrm{SFR}$ to very low values, consistent with $z=0.2$ blank field galaxies.

\begin{figure}
\begin{center}
\includegraphics[width=0.470\textwidth]{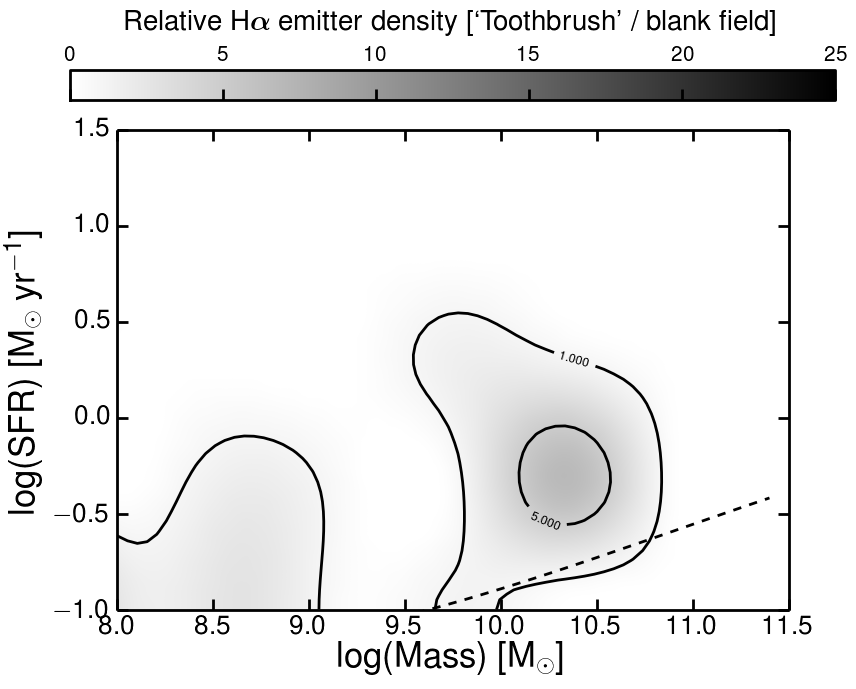}
\end{center}
\caption{Relative density of H$\alpha$ emitters of the `Toothbrush' field compared to a blank field from \citet{2008ApJS..175..128S}, plotted in the stellar mass - star-formation rate plane. The $10\%$ completeness in the SFR measurements is plotted in the dashed line. Note the stark difference with the `Sausage' field. The `Toothbrush' field contains a lower or similar number of emitters as compared to blank fields at all masses and SFRs. The slight enhancement up to a factor of $5$ can be explained by Poissonian errors or cosmic variations. The `Toothbrush' not only follows the SFR-mass relation, but also has the same density of H$\alpha$ emitters as blank fields in each part of the relation.}
\label{fig:toothdiv}
\end{figure}

\subsection{Stellar masses of cluster galaxies}
The cluster galaxies ($10^{8-10.5}$ $M_\odot$) fall on the mass-SFR relation (see Fig.~\ref{fig:sfrmass}). Therefore, even though an exceptionally high-fraction of galaxies are star-forming, given they are located in a cluster, they have SFR typical for their mass. This is in agreement with \citet{2013MNRAS.434..423K} that study the star forming properties of the cluster Cl0939+4713 at $z = 0.4$. Nevertheless, the distribution of H$\alpha$ galaxies within the mass-SFR relation is strikingly different from the field.  To quantify the distribution of H$\alpha$ galaxies in our two clusters in the stellar mass-SFR plane, we compare with the results from \citet{2008ApJS..175..128S}, obtained over volumes $\sim10$ times larger than the survey of the `Sausage' and `Toothbrush' clusters. We bin the H$\alpha$ emitters in mass and SFR and divide the 2D histogram by the different volumes probed by the three surveys. This ensures we obtain comparable results for the clusters and the blank field. We then divide the normalised mass-SFR distribution of emitters for the clusters by the normalised H$\alpha$ distribution for the COSMOS blank field. The bicubicly-interpolated results for the `Sausage' and the `Toothbrush' fields can be visualised in Figs. \ref{fig:sausdiv} and \ref{fig:toothdiv}, respectively. The `Sausage' field contains elevated numbers of H$\alpha$ emitters at all masses and SFRs. The cluster is exceptionally abundant in high-mass ($>10^{10}M_\odot$), highly-star forming ($1-10 M_\odot$ yr$^{-1}$) galaxies. The `Sausage' has a density of H$\alpha$ emitters $20-25$ times higher than a blank field. Note that towards low SFRs, our results suffer from incompleteness so the relative abundance of H$\alpha$ emitters is underestimated. In conclusion, the enhancement in the `Sausage' H$\alpha$ emitter density is seen at all masses, but particularly at the high masses. The `Toothbrush' field H$\alpha$ emitters density is consistent with results obtained from blank fields, with mild over-densities of relatively high-mass, low SFRs H$\alpha$ emitters ($\sim10^{10}M_\odot$, $\sim0.3 M_\odot$ yr$^{-1}$). However, this enhancement in one mass-SFR bin could easily be explained by Poissonian errors and/or cosmic variance.

\subsection{Cause for enhanced star formation in the `Sausage' cluster}\label{sec:discussion:interp}

Both the `Sausage' and the `Toothbrush' clusters have undergone mergers that induced travelling shock waves to propagate through the ICM and accelerate ICM electrons to relativistic speeds. In the case of the `Sausage' cluster, a binary merger lead to the formation of two unequal relics along the merger axis \citep{2011MNRAS.418..230V}. The northern relic has a stronger Mach number than the southern one by a factor of $\sim2$ \citep{2013A&A...555A.110S}. In the `Toothbrush' cluster, simulations and X-ray data indicate a more complicated merger scenario with a main binary merger, possibly followed by a smaller mass interloper participating in the merger towards the end \citep{2012MNRAS.425L..76B,2013MNRAS.433..812O}. The northern relic is also larger than the relic located towards the south-east side of the cluster. Simulations indicate that the `Toothbrush' could also be a more advanced merger than the 'sausage' with a core passage time of the main sub-clusters happening $\sim2$ Gyr ago, compared to $\sim0.5-1$ Gyr ago for the `Sausage' \citep{2011MNRAS.418..230V,2012MNRAS.425L..76B}.

Through our H$\alpha$ observations we discover and confirm numerous star-forming galaxies nearby the relics in the `Sausage' cluster and along the merger axis. Also in the `Toothbrush' cluster there is a clump of star-forming galaxies in the vicinity of the southern shock. The concentration of emitters close to the southern relic in the `Toothbrush' shows that the shock front might have passed through a group of gas-rich galaxies. There is indication that the cluster is a more complicated merger with a third smaller cluster responsible for the straight northern relics and that could also cause the formation of the southern relic, at a later stage than the northern one. If this smaller sub-clump was a low-mass group and still contained numerous gas-rich galaxies that could explain why there is a striking difference between the galaxy populations towards the northern and the southern relic in the `Toothbrush' cluster. 

Another point to note is that the distribution of H$\alpha$ emitters around the clusters is very different. The environment of the `Toothbrush' cluster is at the level expected for the field, while the `Sausage' clusters seems to be embedded in a more SFR-rich region with filamentary H$\alpha$ structures.

The gravitationally decoupled shock travels ahead of the ICM gas, dark matter and galaxies and therefore could have interacted with these components at some point since the sub-clusters merged. In terms of its interaction with galaxies, a shock wave will heat any available intra-galactic gas and increase turbulence and induce instabilities in the gas. The gas can then collapse into clouds which are dense enough to start forming stars. The scenario was proposed by \citet{1989MNRAS.239P...1R} for high redshift galaxies where the powerful radio jet was suggested to induce star-formation along its propagation axis. Radio jets with high Mach numbers ($M\sim10-100$) at their terminating shocks have too much power and entrain the low-entropy gas away from the galaxy uplifting it in buoyant bubbles \citep[e.g.][]{2010MNRAS.407.2063W}. \citet{2010MNRAS.407.2063W} found H$\alpha$ filaments in the immediate downstream area from the jet-termination shock front in M87. \citet{2012MNRAS.421.1603C} and \citet{2014arXiv1409.7700H} found evidence for shock-induced shells around star-forming clumps. The locations of these shells along the optical filaments connecting the two radio jets in the Centaurus A galaxy is consistent with material being swept along the back flow of the jet and ionised through shocks. In the case of the high-redshift galaxy 4C+41.17, \citet{2014arXiv1404.7539S} confirmed the presence of a bow-shock structure inducing star-formation near the core of the galaxy. In the context of interacting galaxies, smooth particle hydrodynamical simulations by \citet{2004MNRAS.350..798B} show that a shock-induced star-formation prescription fits very well observations of NGC 4676. \citet{2009PASJ...61..481S} similarly found in their simulations a vigorous, shock-induced starburst event at the interface of two colliding disk galaxies. By contrast to radio-jet terminating shocks, the merger shocks in the `Sausage' and the `Toothbrush' cluster have Mach numbers of at most $4$. \citet{2014MNRAS.440.3416O} have shown using Chandra data that minor shock fronts are ubiquitous within the ICM of the `Sausage' cluster. 

We speculate that low Mach number, extended shocks might have a higher chance of not removing the gas from the galaxy, but increasing the turbulence. We suggest that there is a tight correlation between the strength of the shock and the ultimate fate of the gas. If the shock is too strong then the gas can be stripped from the host galaxy in the same fashion as it happens when a galaxy is infalling into the cluster potential with relative speeds of $1000-2000$ km s$^{-1}$. For example, there is evidence that the galaxies in the `Sausage' cluster are moving along the merger axis, following the merger direction \citep{2013A&A...555A.110S}. Given the speed at which the collissionless shock front travels with respect to the ICM \citep[$\sim2500$ km s$^{-1}$,][]{modelling} and the relative collision speed ($\sim2250$ km s$^{-1}$, Dawson et al. in preparation), the galaxies seem to be trailing behind the shock front, but are ahead of the ICM, indicating they are possibly moving at a speed of $500-2000$ km s$^{-1}$ with respect to the ICM gas. This is comparable or lower than the speeds reached by galaxies infalling into the cluster via accretion. Note that there is a velocity dispersion of $\sim1000$ km s$^{-1}$, so some galaxies will be going much faster/slower relative to the shock.

The shock-induced SF interpretation is in line with simulations by \citet{roediger2014}. They show that star formation lasting for up to a few $100$ Myr can happen with a delay of $\sim10$ Myr after a passage of the shock. The newly-born stars build up with a tail trailing the shock direction. Multiple episodes of SF can happen after the passage of the shock  and therefore a gradient of ages within these tails or between galaxies located at different distance from the shock is not necessarily expected. Given the delayed start of SF, our H$\alpha$ selection might miss the galaxies very close to the shock fronts, where the star-formation might not have started yet. 

A requirement for the shock and merger to increase the SFR is that the galaxies within the sub-clusters are still relatively gas rich. The H$\alpha$ emitters in the `Sausage' cluster are relatively massive (significant numbers of star-forming galaxies with stellar masses over $10^{10}$ $M_\odot$, see Fig.~\ref{fig:histmass_norm}). The shock is expected to traverse a galaxy within a very short timescale of about $10-50$ Myr. Hence the shock quickly induces turbulence in the gas, after which the gas cools and collapses. The high rate of SF following the shock passage can quickly deplete the gas reservoir. Part of the gas fuels SF, while the rest, is removed from the galaxy through strong outflows. A fraction of the gas located towards the outer disk of the galaxy could be easily stripped by the shock. We therefore expect the passage of the shock to lead to a steep rise in SF for $10-100$ Myr, followed by a quick quenching of the galaxy and a shut-down in the formation of new stars. 

As shown in Sobral et al (in prep), there is evidence for strong outflows in some of the `Sausage' cluster galaxies with asymmetric and P-cygni profiles in the H$\alpha$, [N{\sc ii}] and [S{\sc ii}] lines and broad components for the [N{\sc ii}] and [S{\sc ii}] forbidden lines. This suggests that even though galaxies are highly star-forming as we observe them, they will quickly evolve into gas-poor galaxies. Given the galaxies reside in a very massive cluster, any outflowing material will easily escape the host galaxy. The high mass-loss rate caused by these outflows leads to a quick depletion of the gas reservoir effectively shutting down star-formation. Hence, the outflows indicate that H$\alpha$ galaxies will be transformed into passive galaxies within a short time scale. We can make an estimate of the rate at which this can happen by comparing the two clusters. The `Toothbrush' cluster hosts about $60$ per cent ($34$) fewer H$\alpha$ galaxies than the `Sausage' cluster. If the time since core passage is $2$ Gyr for the `Toothbrush' compared to $1$ Gyr for the `Sausage' cluster, SF in these galaxies is being shut-down at a rate of $34$ galaxies per Gyr, or one galaxy every $\sim30$ Myr. If no additional accretion of fresh matter happens, the `Toothbrush' will have a complete shut-down of star formation within $1$ Gyr of when we observe it. `Right now', the `Toothbrush' galaxy population is fully passively-evolving. For the `Sausage', the process of star-formation shut down will take another $2$ Gyr. Given that the stellar mass of the `Sausage' galaxies is $10^{8-10.5}$ $M_\odot$, and assuming a molecular gas $M_\mathrm{gas}$ to gas plus stellar mass $M_\mathrm{gas}+M_{\star}$ ratio of $0.1$ \citep{1991ARA&A..29..581Y}, our galaxies have about $10^{7-9.5}$ $M_\odot$ molecular gas. The molecular gas has two possible fates: it either contributes to star-formation or it is removed from the host galaxy, either via outflows or ram pressure stripping. The characteristic H$\alpha$ luminosity of the `Sausage' galaxies of $10^{41.6}$ erg s$^{-1}$ is equivalent to a rate of $\sim3$ $M_{\odot}$ yr$^{-1}$ of molecular gas being converted into new stars. If all the molecular gas is consumed through SF, the low mass galaxies would use up their molecular gas within $1$ Myr, while the galaxies located at the high mass end would require $1$ Gyr. Assuming a maximal mass loss through outflows equal to the rate of conversion to SF, star formation in the `Sausage' galaxies would last up to $0.5$ Gyr. 

Another calculation we can make using the difference of age between the clusters is the quantity of stellar mass being added to the cluster volume. We assume that the two clusters are relatively similar in their properties apart from the `Toothbrush' being an older merger compare to the `Sausage'. In this scenario, the `Toothbrush' represents a "look into the future" of a `Sausage'-like cluster in $0.5$ Gyr. The `Sausage' cluster $\rho_\mathrm{SFR}$ is $\sim6.9$ $M_{\odot}$ yr$^{-1}$ Mpc$^{-3}$ (corrected for the volume occupied by the cluster), while  the `Toothbrush' is only $\sim0.3$ $M_{\odot}$ yr$^{-1}$ Mpc$^{-3}$, a drop which could happen, given the simulations of the two clusters, over $0.5$ Gyr. Within every Mpc$^3$, a mass of $\sim1.5\times10^9$ $M_\odot$ of stars would be formed. Given the entire cluster volume is expected to be about $30$ Mpc$^3$, the total stellar mass added to the cluster is $\sim4.5\times10^{10}$ $M_\odot$, less than $3$ per cent of the mass of the Milky Way. A similar mass is lost through outflows. Therefore, even though many galaxies within merging clusters can go through episodes of vigorous star-formation, once enough time passes, this will not necessarily reflect in the total mass of passive galaxies.

As an interpretation for our results, we suggest that even in a hot cluster atmosphere, as long as galaxies retain some of their gas content until the passage of the merger shock, it is possible to observe high levels of star formation in a large fraction of cluster galaxies. The influence of shocks seems to supersede the passive evolution of galaxies within the cluster environment and prevent the rapid loss of gas through interactions with the ICM. Rather, the shock favours the retainment of gas within its host galaxy. This is consistent with the finding that the number of high-mass star-forming galaxies is especially boosted in the `Sausage' cluster. High stellar-mass galaxies reside in massive dark matter haloes with a strong gravitational pull capable of holding the gas. Shocks probably induce instabilities in the gas which collapses into star-forming clouds and also increase AGN activity that produces large mass outflows. Therefore, while momentarily the galaxies close to the shock front will exhibit high levels of H$\alpha$ emission, the fast consumption of their gas will lead to an accelerated evolution from gas-rich to gas-poor ellipticals or S0s, compared to other cluster galaxies allowed to passively evolve. 

\section{CONCLUSIONS}\label{sec:conclusion}

We conducted an H$\alpha$ survey of two post-core passage, merging clusters which host Mpc-wide travelling shock waves (the `Sausage' and the `toothbrush clusters). Using optical broad and narrow band data and spectroscopy, we are able to draw a number of conclusions.
\begin{itemize}
\item We robustly select line emitters towards the two clusters using custom-made narrow-band filters. Down to a similar equivalent width, but different luminosity limits, we select $201$ and $463$ line emitters towards the `Sausage' and the `Toothbrush' field of view, respectively. We separate between H$\alpha$ emitters at the cluster redshift and other higher-redshift emitters using colour-colour diagnostics, photometric and spectroscopic data. 
\item Based on photometric redshifts, the emitter population for the `Sausage' cluster is clearly dominated by H$\alpha$ emitters with a fraction of $62$ per cent, $52.5$ per cent of which are confirmed by spectroscopy. The bulk of the emitters are located in the cluster. 
\item In the case of the `Toothbrush', $89$ per cent of the emitters are not H$\alpha$. The bulk of the H$\alpha$ emitters are located in the field environment around the cluster.
\item We find a clear enhancement in the number density of H$\alpha$ emitters in the `Sausage' compared to a blank field $\log \phi^*$ ($-1.37\pm0.04$ versus $-2.65^{+0.27}_{-0.38}$), pointing towards a very high fraction of the cluster population being star-forming. This is a highly surprising result given the high X-ray temperature of the ICM ($9-13$ keV). The results also hold if we look only at emitters around the shock fronts or within the cluster volume (as defined from weak lensing data). The star-formation rate density for the cluster is at the order of $15$ times the peak of the star-formation history of the Universe. The average star-formation rate of galaxies along the merger axis is higher than away from the shock fronts.  The cluster has a density of high-mass, highly star-forming galaxies $20-25$ higher than blank fields.
\item The normalisation $\log \phi^*$ of the H$\alpha$ luminosity function in the `Toothbrush' cluster is $-1.76\pm0.02$, compared to $-2.65^{+0.27}_{-0.38}$ for a blank field. Owing to a specific luminosity $\log L^*=40.75\pm0.01$ in the cluster lower than the field ($41.57^{0.38}_{-0.23}$), the `Toothbrush' overall star-formation rate density is consistent with blank fields at $z=0.2$.
\item We find that the relation between the SFR and stellar mass for the two clusters is very similar to that of blank fields. However, the density of H$\alpha$ emitters in the `Sausage' cluster is boosted compared to the field, especially in the high mass, high SFR regime, where the cluster is $20-25$ times denser than the COSMOS field.
\item Accounting for the different ages of the two clusters, we measure a rate of $1$ star-forming galaxy being transformed into a non-star forming galaxy every $\sim$ 30 Myr to transform a star-formation rich, `Sausage'-like cluster into a `Toothbrush'-like cluster devoid of star-forming galaxies.
\item We interpret our results as shock-induced star-formation. In line with simulations and other observational results, the merger and shock waves lead to a momentary increase in star-formation in gas rich galaxies. This is turn accelerates the turn-off of star-formation owing to a rapid consumption of the molecular gas supply. This effects seems to be happening at all masses and star-formation rates, but predominantly in the high-mass regime.
\end{itemize}

\section*{Acknowledgements}
We thank the anonymous referee for the his/her comments that helped improving the clarity of the paper. We thank Florian Pranger, David Carton, Francois Mernier, Monica Turner and Mattia Fumagalli for useful discussions. Based on observations made with the Isaac Newton Telescope (proposals I12BN003 and I13BN006) and the William Herschel Telescope (proposal W13BN006, W14AN012) operated on the island of La Palma by the Isaac Newton Group in the Spanish Observatorio del Roque de los Muchachos of the Instituto de Astrof{\'i}sica de Canarias. Based in part on data collected at Subaru Telescope, which is operated by the National Astronomical Observatory of Japan. Also based on observations obtained through the OPTICON programme 13B055 with MegaPrime/MegaCam through, a joint project of CFHT and CEA/DAPNIA, at the Canada-France-Hawaii Telescope (CFHT) which is operated by the National Research Council (NRC) of Canada, the Institute National des Sciences de l'Univers of the Centre National de la Recherche Scientifique of France, and the University of Hawaii. The research leading to these results has received funding from the European Community's Seventh Framework Programme (FP7/2007-2013 and FP7/2013-2016) under grant agreements numbers RG226604 and 312430 (OPTICON). Some of the data presented herein were obtained at the W.M. Keck Observatory, which is operated as a scientific partnership among the California Institute of Technology, the University of California and the National Aeronautics and Space Administration. The Observatory was made possible by the generous financial support of the W.M. Keck Foundation. This research has made use of the NASA/IPAC Extragalactic Database (NED) which is operated by the Jet Propulsion Laboratory, California Institute of Technology, under contract with the National Aeronautics and Space Administration. This research has made use of NASA's Astrophysics Data System. AS acknowledges financial support from the Netherlands Organisation for Scientific Research (NWO). DS acknowledges financial support from NWO through a VENI fellowship, from FCT through an FCT Investigator Starting Grant ad Start-up Grant (IF/01154/2012/CP0189/CT0010) and from FCT grant PEst-OE/FIS/UI2751/2014. RJvW acknowledges support provided by NASA through the Einstein Postdoctoral grant number PF2-130104 awarded by the Chandra X-ray Center, which is operated by the Smithsonian Astrophysical Observatory for NASA under contract NAS8-03060. Part of this work performed under the auspices of the U.S. DOE by LLNL under Contract DE-AC52-07NA27344.

\bibliographystyle{mn2e.bst}
\bibliography{Halpha2014}

\appendix
\section{Dust extinction screens}
\begin{figure*}
\begin{center}
\includegraphics[trim=0cm 0cm 0cm 0cm, width=0.33\textwidth]{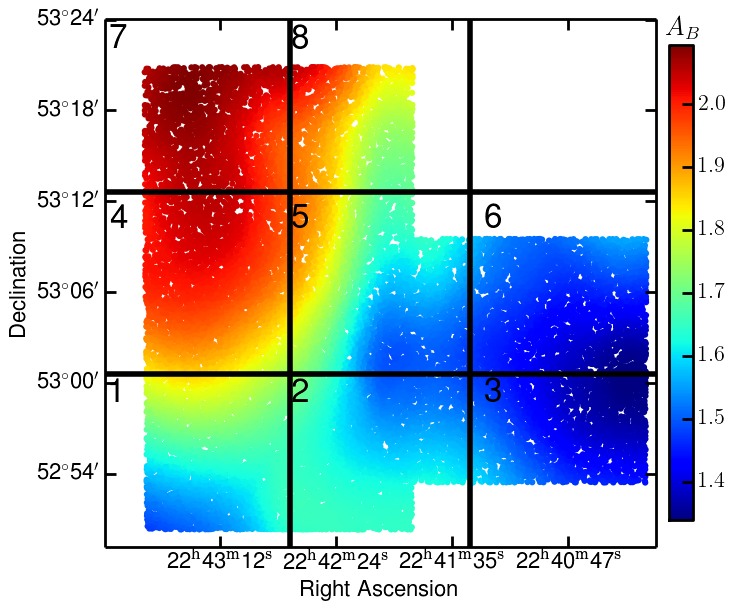}
\includegraphics[trim=0cm 0cm 0cm 0cm, width=0.33\textwidth]{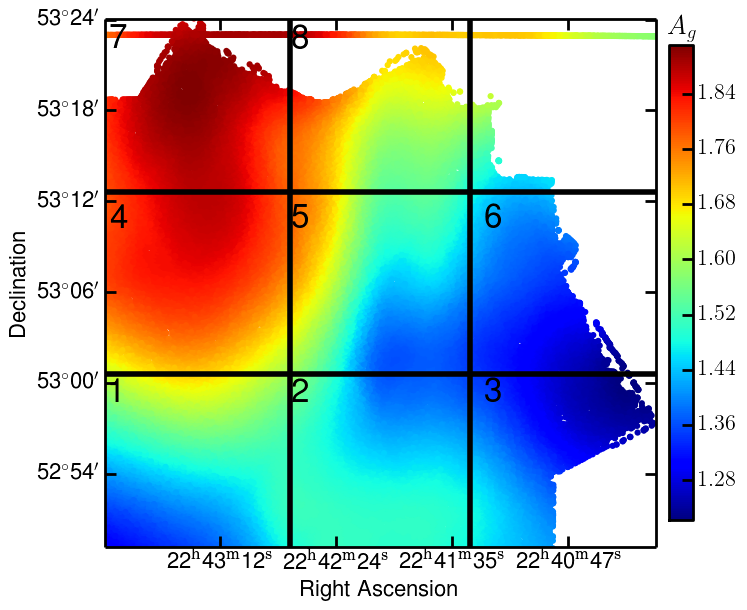}
\includegraphics[trim=0cm 0cm 0cm 0cm, width=0.33\textwidth]{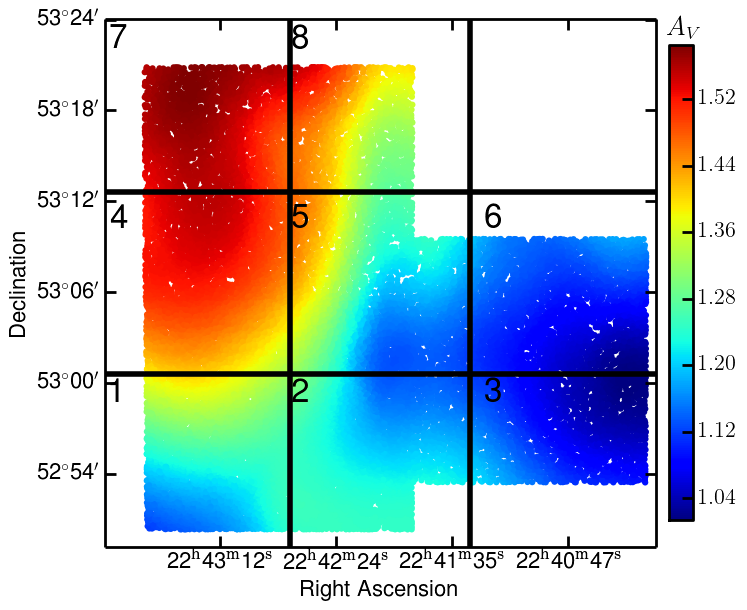}\\
\includegraphics[trim=0cm 0cm 0cm 0cm, width=0.33\textwidth]{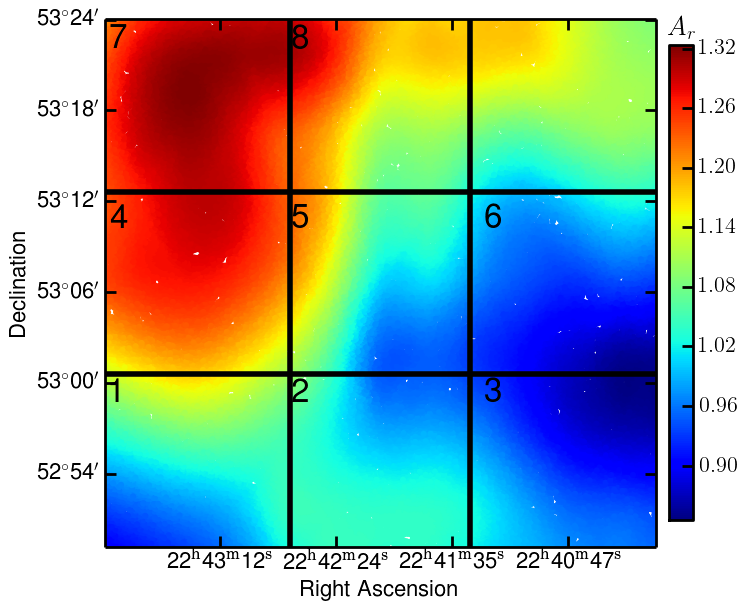}
\includegraphics[trim=0cm 0cm 0cm 0cm, width=0.33\textwidth]{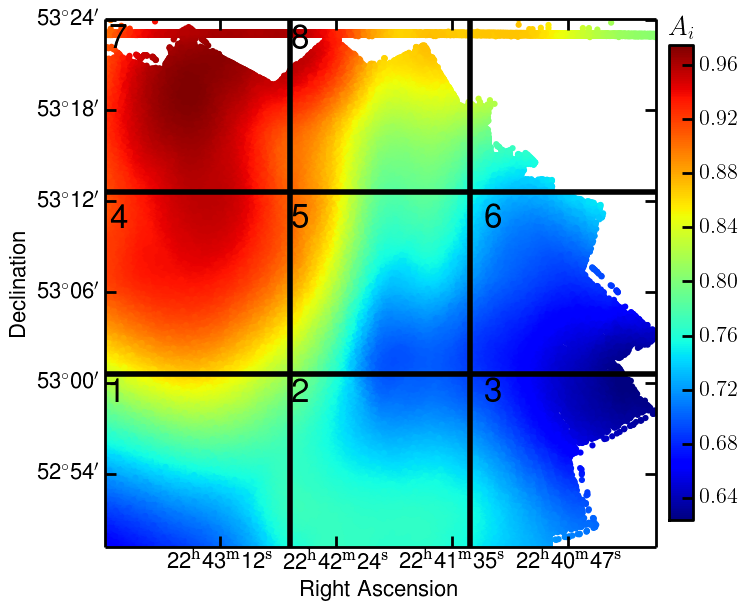}
\includegraphics[trim=0cm 0cm 0cm 0cm, width=0.33\textwidth]{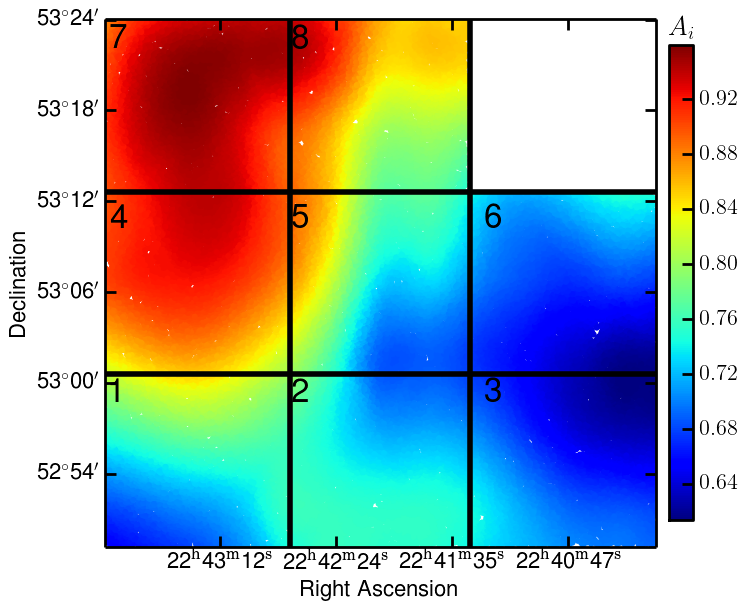}\\
\includegraphics[trim=0cm 0cm 0cm 0cm, width=0.33\textwidth]{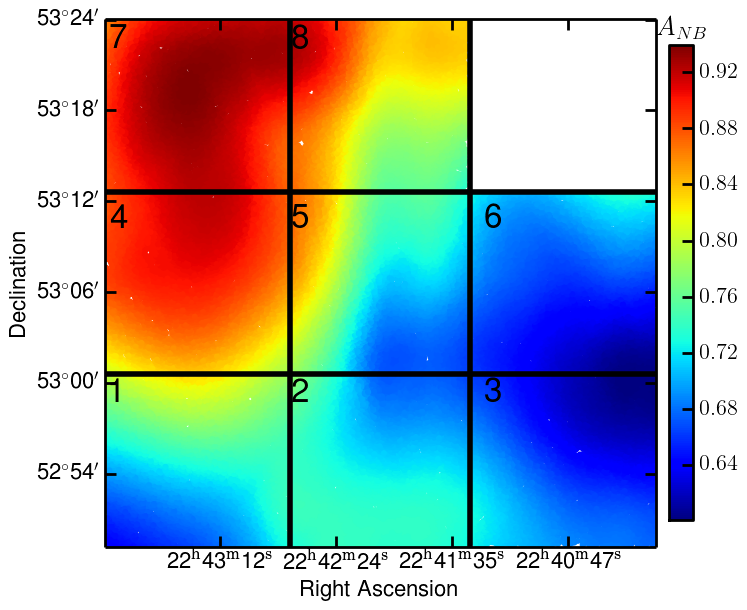}
\includegraphics[trim=0cm 0cm 0cm 0cm, width=0.33\textwidth]{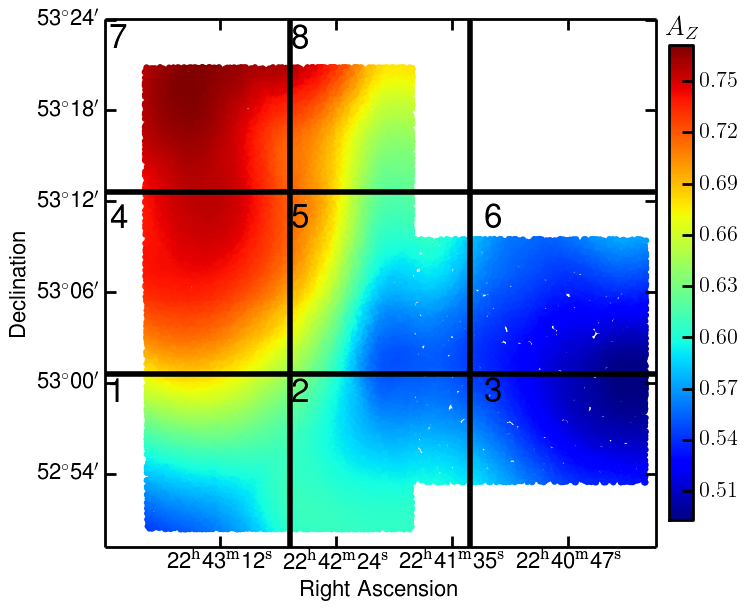}
\includegraphics[trim=0cm 0cm 0cm 0cm, width=0.33\textwidth]{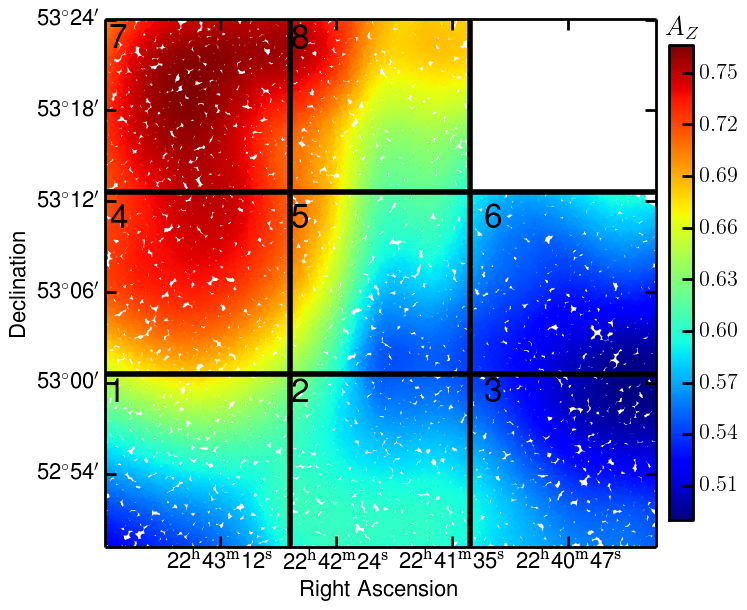}\\
\caption{`Dust screens' for the `Sausage' for the different filters from top-left to bottom right: WHT B, WHT V, Subaru g, CFHT r, Subaru i, INT i, INT NB, WHT z, INT z. The figures are all on the same RA \& DEC scale and serve to display the FOV coverage with each camera. The values of the dust attenuation are calculated for each source in the FOV and all sources are plotted as individual points. All filters cover the full extent of the cluster. Note the NB extinction is effectively the same as for the i band filter, as the NB filter is centred close to the central wavelength of the i band filter. The dust extinction, measured in magnitudes, is based on measurements from \citet{2011ApJ...737..103S}. Note the different scales of the figures: the dust extinction and its variation across the FOV increase significantly towards the blue side of the optical spectrum. The dust extinction variations across the FOV lead to differences in depth across the FOV, therefore the FOV has been divided into eight areas for completeness study purposes.}
\label{fig:dust:sausage}
\end{center}
\end{figure*}

\begin{figure*}
\begin{center}
\includegraphics[trim=0cm 0cm 0cm 0cm, width=0.33\textwidth]{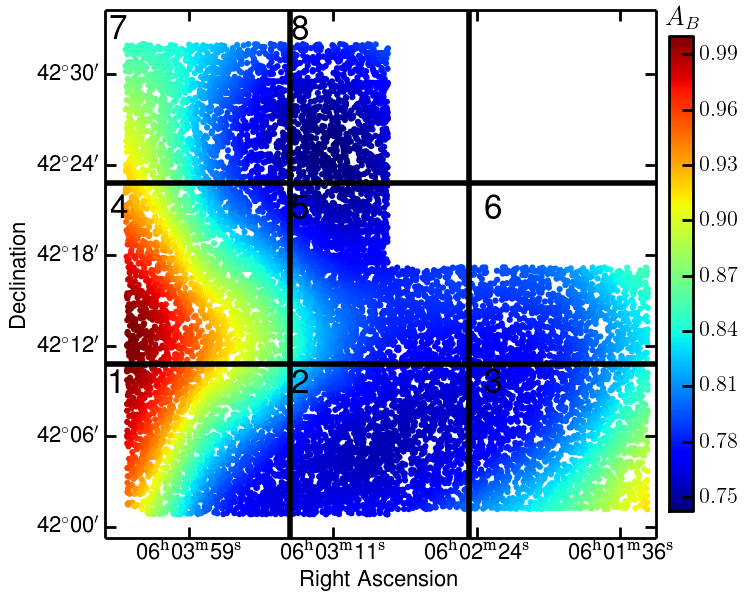}
\includegraphics[trim=0cm 0cm 0cm 0cm, width=0.33\textwidth]{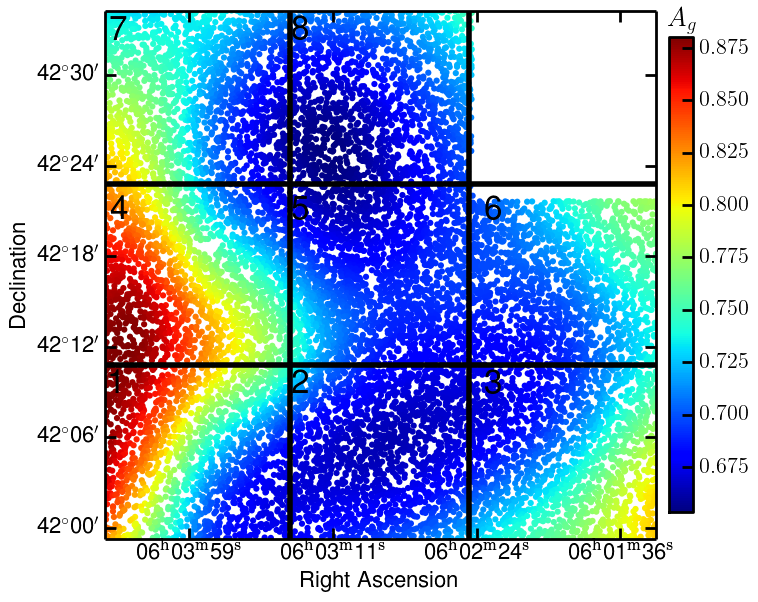}
\includegraphics[trim=0cm 0cm 0cm 0cm, width=0.33\textwidth]{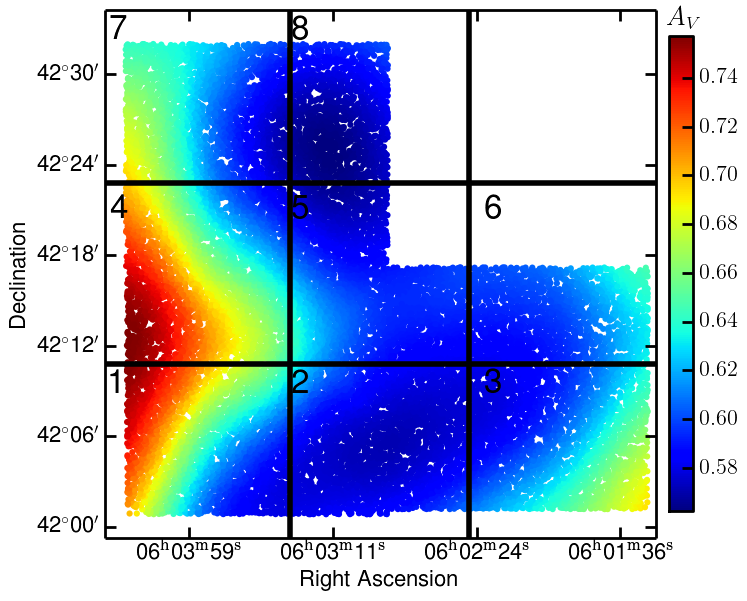}\\
\includegraphics[trim=0cm 0cm 0cm 0cm, width=0.33\textwidth]{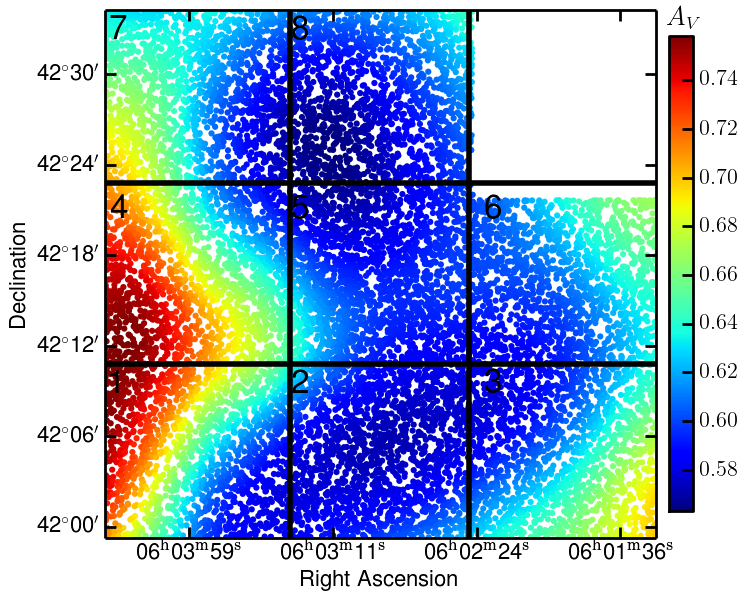}
\includegraphics[trim=0cm 0cm 0cm 0cm, width=0.33\textwidth]{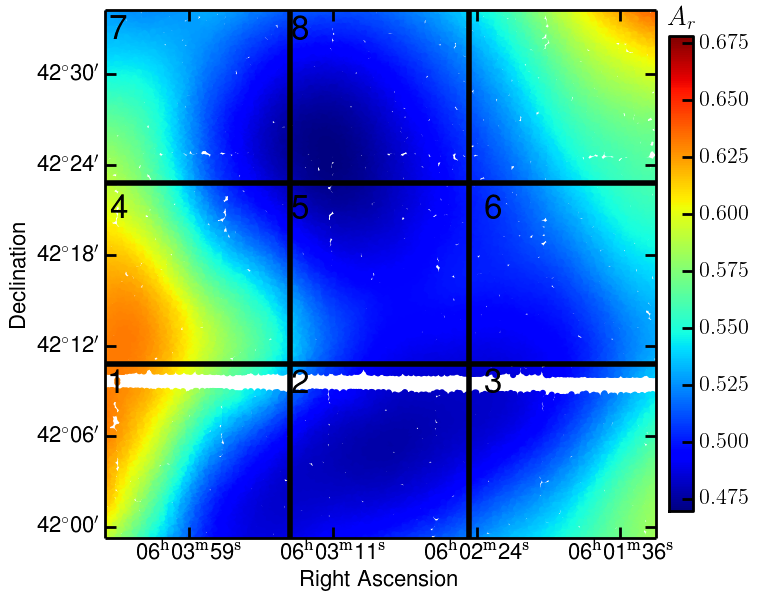}
\includegraphics[trim=0cm 0cm 0cm 0cm, width=0.33\textwidth]{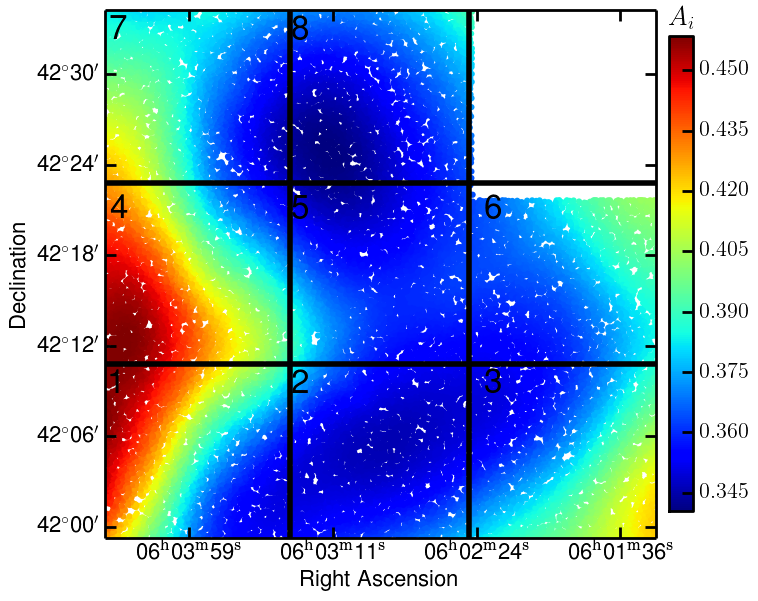}\\
\includegraphics[trim=0cm 0cm 0cm 0cm, width=0.33\textwidth]{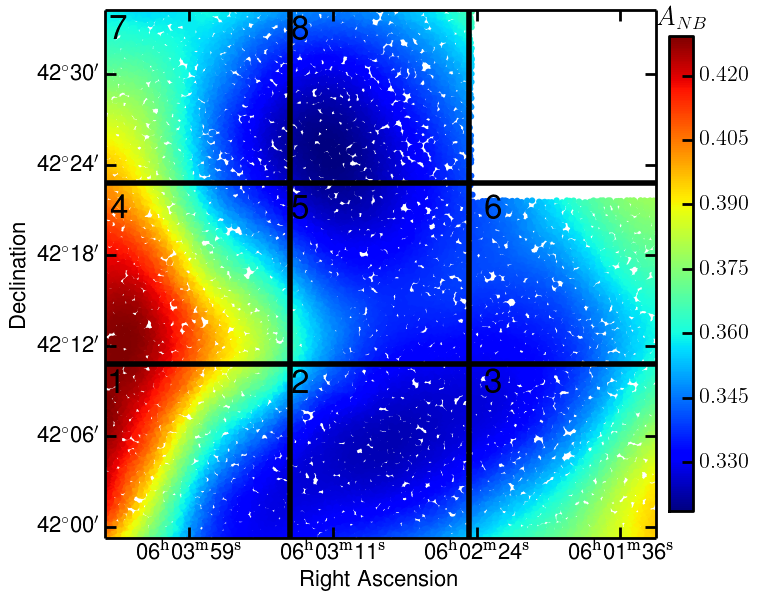}
\includegraphics[trim=0cm 0cm 0cm 0cm, width=0.33\textwidth]{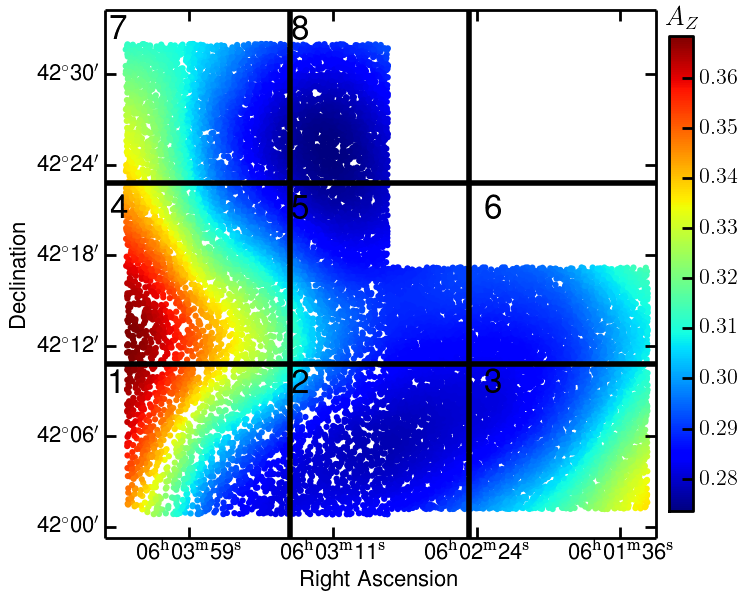}
\includegraphics[trim=0cm 0cm 0cm 0cm, width=0.33\textwidth]{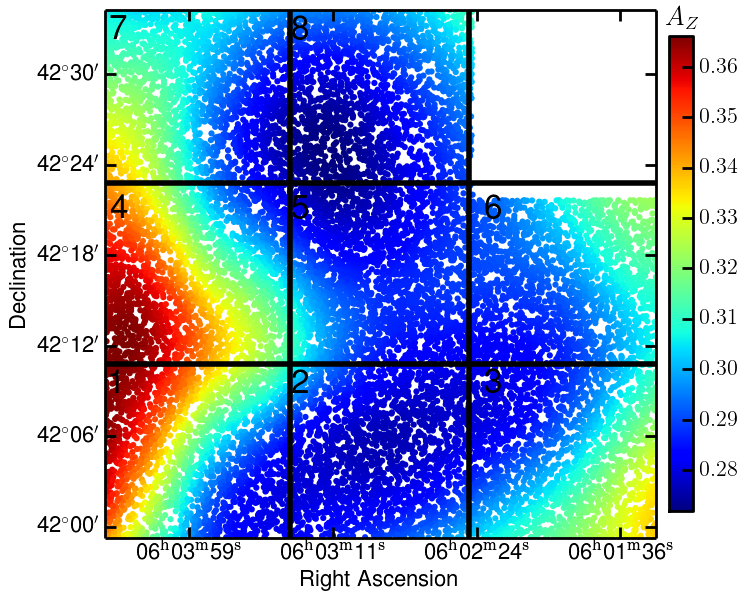}
\caption{The same as Fig.~\ref{fig:dust:sausage}, but for the `Toothbrush' field. The figures are in order from top-left to bottom right: WHT B, INT g, WHT V, INT V, CFHT r, INT i, INT NB, WHT z, INT z. Note that because the dust attenuation is plotted for individual sources, the lower source density than the `Sausage' field is apparent.}
\label{fig:dust:toothbrush}
\end{center}
\end{figure*}

\end{document}